\documentclass[sn-aps]{sn-jnl}

\usepackage{graphicx}%

\usepackage{pgfplots}
\usepackage{multirow}%
\usepackage{amsmath,amssymb,amsfonts}%
\usepackage{amsthm}%
\usepackage{mathrsfs}%
\usepackage[title]{appendix}%
\usepackage{xcolor}%
\usepackage{textcomp}%
\usepackage{manyfoot}%
\usepackage{booktabs}%
\usepackage{algorithm}%
\usepackage{algorithmicx}%
\usepackage{algpseudocode}%
\usepackage{listings}%
\usepackage{float}
\usepackage{graphicx}
\usepackage{tikz}
\usetikzlibrary{shapes.geometric, arrows, positioning}

\begin{document}

\title[Article Title]{A Survey of Machine Learning Techniques for Improving Global Navigation Satellite Systems}


\author*[]{\fnm{Adyasha}\sur{Mohanty and Grace Gao*}}\email{gracegao@stanford.edu}


\affil[]{\orgdiv{Aeronautics and Astronautics}, \orgname{Stanford University}, \orgaddress{\street{496 Lomita Mall}, \city{Palo Alto}, \postcode{94305}, \state{CA}, \country{USA}}}


\abstract{
Global Navigation Satellite Systems (GNSS)-based positioning plays a crucial role in various applications, including navigation, transportation, logistics, mapping, and emergency services. 
Traditional GNSS positioning methods are model-based and they utilize satellite geometry and the known properties of satellite signals.
However, model-based methods have limitations in challenging environments and often lack adaptability to uncertain noise models.
This paper highlights recent advances in Machine Learning (ML) and its potential to address these limitations. It covers a broad range of ML methods, including supervised learning, unsupervised learning, deep learning, and hybrid approaches.
The survey provides insights into positioning applications related to GNSS such as signal analysis, anomaly detection, multi-sensor integration, prediction, and accuracy enhancement using ML. 
It discusses the strengths, limitations, and challenges of current ML-based approaches for GNSS positioning, providing a comprehensive overview of the field.}




\maketitle

\section{Introduction}\label{sec1}

Global Navigation Satellite Systems (GNSS)-based positioning underpins numerous essential applications, enabling efficiency, safety, and reliability across various industries.
It serves a wide range of applications, including navigation, transportation, logistics, mapping, surveying, and precision agriculture, among others. 
Additionally, emergency services rely on GNSS for search and rescue operations. 
Maritime navigation, and aviation also heavily rely on GNSS for positioning information that enhances situational awareness and reduces response times. 
Furthermore, GNSS plays a crucial role in synchronizing critical infrastructure systems such as power grids, telecommunication networks, and financial transactions~\cite{misra2006global, gebre2009gnss}.


However, GNSS measurements are subject to various sources of error that can affect positioning accuracy~\cite{karaim2018gnss, groves2015gnss, kuusniemi_gnss_2004}. 
One source of error is signal interference that is caused by natural or man-made obstructions, such as tall buildings or dense foliage, leading to signal blockage, Non-Line-of-Sight (NLOS) errors, and multipath (MP) effects in urban environments. 
Another factor is atmospheric delays caused by the ionosphere and troposphere, which can influence the speed of the signals and introduce errors in distance measurements. 
Additionally, clock inaccuracies in both the satellites and receivers can contribute to errors in timing and positioning calculations. 
Other sources of error include satellite orbit inaccuracies and receiver noise.
Mitigating these error sources is crucial in improving GNSS positioning performance for various applications.

Traditionally, model-based methods are used for GNSS positioning and error mitigation/detection because of the following advantages.
Model-based methods incorporate knowledge about signal propagation characteristics in urban environments via statistical models that capture the characteristics of GNSS signals in urban environments. 
These models are based on well-understood physical principles, which have been refined and validated over decades, making their behavior predictable in different environments. 
Model-based algorithms are also less computationally intensive and do not necessarily need vast amounts of labeled data for training.

Model-based methods for GNSS positioning include Newton-Raphson~\cite{nadler1998efficient}, Weighted Least Squares (WLS)~\cite{misra2006global}, and Kalman Filters~\cite{kalman1960new}.
While the Newton-Raphson method enables iterative refinement of the receiver's position estimate~\cite{nadler1998efficient, nadler2000algorithms}, WLS statistically optimizes the solution by assigning weights to each observation based on the measurement quality~\cite{dailey}.
Kalman Filters estimate the state recursively by combining measurements with known system dynamics~\cite{kalman1960new}. 
Other techniques include differential positioning which uses measurements from both the receiver as well as a reference station to correct for common errors affecting both the reference and receiver, such as atmospheric delays, clock errors, and orbit inaccuracies~\cite{walsh1997real, misra2006global}. 
Real-time kinematic (RTK) is a commonly used differential positioning technique in applications such as surveying and precision agriculture~\cite{wang1999stochastic, eissfeller2001real}. 
It involves the use of a base station with known coordinates and a rover receiver. The base station provides correction data to the rover in real-time, allowing for centimeter-level positioning accuracy. 
Similarly, another technique, notably, Precise Point Positioning (PPP) can achieve centimeter-level accuracy without external reference stations~\cite{hofmann2007gnss}. 
It utilizes precise satellite orbit and clock information, along with correction models for atmospheric delays. 
Differential positioning techniques, such as RTK and PPP, often rely on the availability of reference stations or precise orbit and clock data. This dependency can limit their practicality and flexibility in remote or challenging environments~\cite{li2022review}. 
Some methods like PPP involve computationally intensive operations and require longer observation times for accurate results. Real-time processing of high-precision positioning can be challenging, particularly in time-critical applications.

Traditionally, errors such as NLOS errors are identified and mitigated using the signal-to-noise ratio (SNR), weighting models, statistical approaches, and, consistency checking~\cite{groves2015gnss, zhu2018gnss}.
MP errors are handled by using elevation-enhanced maps, successive-time double differences~\cite{elevation}, and analysis of the SNR fluctuation~\cite{tokura2014using}, among others. 
Receiver clock errors are typically mitigated using a clock-steering mechanism~\cite{rcvrerror}, differencing between satellites and estimating the error as an additional unknown parameter in the position estimation process~\cite{Chen2015Estimation}.
Signal propagation errors, such as ionospheric and tropospheric errors are removed using dual-frequency receivers~\cite{bassiri1993higher} and with models such as the Klobuchar~\cite{klobuchar1987ionospheric} and  Saastamoinen models~\cite{saastamoinen1972atmospheric}. 
Satellite orbital errors are mitigated using a global or local network of corrections for the satellite positions or in a post-processed manner~\cite{han2001accurate, langley1991orbits}. 

While model-based methods are extensively used for positioning, error detection, and mitigation, they have certain limitations.
Model-based techniques face challenges
due to their strict initial assumptions concerning sensor noise and model parameters.
The conventional model-based methods often assume noise to be Gaussian (or normally distributed), which simplifies the mathematics involved in filtering and estimation processes, such as in the application of Kalman filters for real-time positioning. 
In real-world scenarios, the noise affecting GNSS signals can deviate significantly from Gaussian behavior. Sources such as MP effects, where signals bounce off surfaces before reaching the receiver, create a complex error structure that is not well-modeled by a normal distribution. Similarly, atmospheric disturbances, signal reflection, and interference can introduce noise with heavy tails or skewed distributions that Gaussian models fail to capture accurately.
Noise characteristics can vary with location, time, and environmental conditions, introducing further complexity. For instance, urban environments might experience more significant MP effects due to tall buildings, while rural areas might have different noise profiles. Temporal changes, such as atmospheric conditions can also affect noise characteristics over time.
Such assumptions limit the adaptability of model-based techniques, especially in challenging environments where the noise characteristics, model parameters, and error models may not adhere to the predefined assumptions~\cite{model_bad, model_bad2, zhao2023fusing, yozevitch_robust_2016}. 
In contrast, ML techniques have emerged as novel approaches in GNSS-based positioning, addressing the limitations of model-based methods. 
These techniques are more suitable for handling nonlinear relationships between variables, can learn from large amounts of data, and adapt to new and changing environments.
ML algorithms can learn hidden and nonlinear relationships from data directly without relying on noise assumptions. 
These algorithms are also robust to missing
data and handle outliers more effectively than model-based methods~\cite{muller_future_2016, kunze_artificial_2018, fayyad_deep_2020, shahbazian_machine_2023}.

Given the significance of ML techniques in enhancing GNSS positioning and performance, there is a need to design a comprehensive survey paper to consolidate and disseminate knowledge in this field. 
In this regard, Jagiwala et al. \cite{jagiwala_possibilities_2022} provide an insightful review, emphasizing the role of support vector machines (SVMs) and convolutional neural networks (CNNs) in enhancing position accuracy. 
While a systematic review of machine learning techniques for GNSS use cases is covered in \cite{siemuri_machine_2021}, our survey paper distinguishes itself by making the following contributions.

\begin{itemize}
    \item  It provides a comprehensive review of a wide range of ML methods applied to GNSS positioning, including supervised learning, unsupervised learning, deep learning, and hybrid approaches. This provides a broader perspective on the subject by showcasing the diverse applications of ML in the field.
    \item It includes the latest research developments and advancements post-2021 in ML techniques for GNSS positioning. This equips readers with a current understanding of recent trends, innovations, and the state-of-the-art in the domain.
    \item Beyond the performance evaluation of machine learning techniques, the paper describes various ML use cases in GNSS. Key topics include using machine learning for signal analysis, anomaly detection, multi-sensor integration, prediction, forecasting, and more.
    \item By evaluating the strengths, challenges, and potential limitations of existing ML techniques, the paper provides readers with an improved understanding of the potential and constraints of ML in enhancing GNSS positioning accuracy.

\end{itemize}

The paper is organized as follows. Section 2 provides a brief background on the relevance of ML methods to GNSS positioning. Section 3 discusses various ML methods for the analysis and classification of GNSS signals, including supervised machine learning techniques such as SVM and decision trees, unsupervised ML methods, deep learning techniques, and hybrid approaches. Section 4 focuses on ML techniques for environmental context and scenario recognition using GNSS measurements, while Section 5 explores ML techniques for anomaly detection and quality assessment. Section 6 covers ML methods for GNSS-based multi-sensor integration, and Section 7 discusses prediction and forecasting techniques leveraging GNSS measurements and AI. In Section 8, techniques for enhancing positioning accuracy and position error modeling are discussed. Section 9 highlights other notable applications of using ML for improving GNSS. Section 10 addresses the limitations and challenges associated with the discussed ML methods. Finally, Section 11 identifies potential areas for future research and development in the field of AI-based GNSS positioning.

\section{Background of ML Methods}

\subsection{Regression methods}
\label{regression}

Regression methods predict continuous numerical values by mapping input features to a target variable using a mathematical model with adjustable parameters. 
The model is trained by minimizing the difference between predicted and actual values. 
We discuss commonly used regression techniques below.

\begin{figure}[h]
   \centering
   \includegraphics[width=0.5\textwidth]{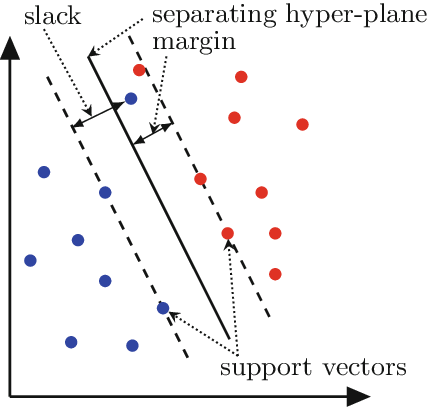}
   \caption{Illustration of SVM from \cite{figsvm}. SVM is a supervised learning algorithm that finds the hyperplane that best separates different classes with the maximum margin. It uses support vectors and kernels to optimize the separation boundary in both linear and non-linear classification tasks.}
   \label{fig:svm}
\end{figure}

\begin{itemize}
    \item Quantile Regression, an extension of traditional regression analysis~\cite{koenker2005quantile}, estimates different quantiles of the target variable's conditional distribution. Unlike ordinary least squares regression, which focuses on the conditional mean, Quantile Regression provides a comprehensive understanding of the conditional distribution by considering multiple quantiles. It achieves this by minimizing a loss function that measures the discrepancy between predicted and actual quantiles. This optimization process determines the optimal parameters governing the relationship between input features and the target variable's quantiles.

    \item     SVMs are used for both classification and regression tasks~\cite{svmcortes1995support}. They identify optimal hyperplanes to separate classes and handle nonlinear data through the kernel trick. Support Vector Regression (SVR), a variation of SVM, fits data by allowing a margin for error and utilizes kernel functions to capture linear and nonlinear relationships~\cite{smola2004tutorial}. An example SVM is shown in Figure~\ref{fig:svm}. 
    Support vectors, identified during training, play a vital role in generalization and prediction. SVR estimates numerical values for new data points by applying learned parameters and support vectors.

\end{itemize}

\subsection{Unsupervised learning methods}
\label{unsupervised}

Unsupervised learning involves training models on unlabeled data without explicit guidance or predefined labels. Instead of predicting specific outcomes, unsupervised learning algorithms focus on discovering hidden patterns, structures, or relationships within the data.
The primary categories of unsupervised learning methods include the following.

\begin{itemize}

\item K-means algorithm is a widely used method for clustering, which partitions a dataset into k distinct, non-overlapping subsets or clusters~\cite{MacQueen1967}. The algorithm assigns each data point to the cluster with the nearest mean, serving as a prototype of the cluster. This process iteratively adjusts the positions of the centroids (the means of the clusters) and reassigns the data points to their closest centroids until the positions of the centroids stabilize, indicating that the clusters are as compact and distinct from each other as possible. 

    \item Autoencoders are neural networks that are commonly used for dimensionality reduction~\cite{hinton2006reducing}. As illustrated in Figure~\ref{fig:auto}, they consist of an encoder network that maps the input data to a lower-dimensional representation, and a decoder network that reconstructs the original input from this representation. Autoencoders learn a compressed and efficient representation of the input data, capturing essential features.

\begin{figure}[h]
   \centering
   \includegraphics[width=0.7\textwidth]{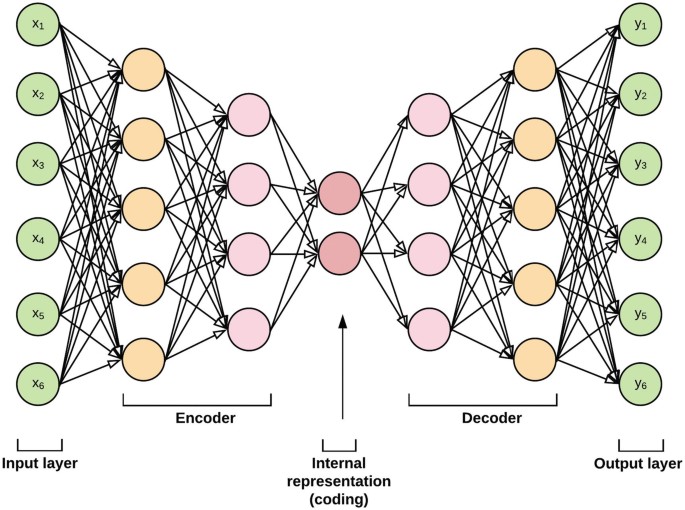}
   \caption{Illustration of an autoencoder from \cite{figauto}. Autoencoder compresses the inputs into a latent-space representation and then reconstructs the output from this representation, aiming to match the original input.}
   \label{fig:auto}
\end{figure}

\item Variational autoencoders (VAE) learn a lower-dimensional latent space representation of input data, capturing its underlying structure and distribution~\cite{autokingma2013}. VAEs consist of an encoder network and a decoder network. The encoder maps input data to a latent space, typically represented by the mean and the variance of a Gaussian distribution. The decoder reconstructs input data from latent space samples. Training VAEs involves optimizing two objectives: reconstruction loss and the Kullback-Leibler (KL) divergence regularization term. The reconstructed output resembles the original input, while the regularization term encourages a structured latent space. VAEs can generate new samples resembling training data and compress data by encoding and decoding it from the latent space.
\end{itemize}

\subsection{Classification methods}
\label{classification}
AI algorithms for classification utilize machine learning techniques to automatically assign data instances into predefined categories or classes based on their features or attributes. These algorithms learn from labeled training data to build models that can accurately classify new, unseen data. 
Two common classification approaches are Decision Trees and Naive Bayes. 

\begin{figure}[h]
   \centering
   \includegraphics[width=0.5\textwidth]{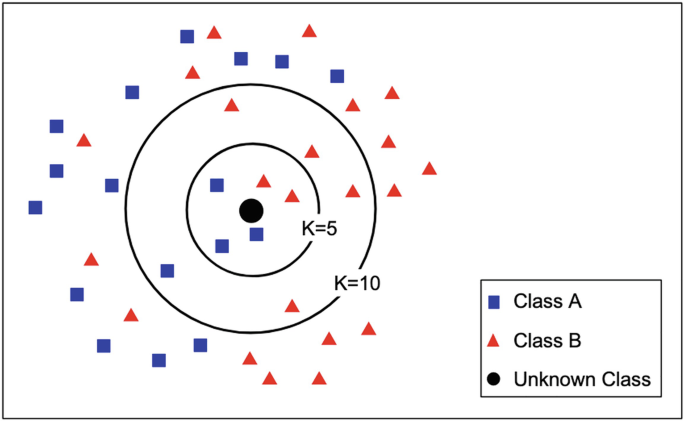}
   \caption{An example illustrating how KNNs are used in classification tasks~\cite{figknn}. KNN is a non-parametric learning algorithm that classifies new cases based on the majority vote of the $k$ most similar instances from the training data, often using distance metrics like Euclidean distance to determine similarity. }
   \label{fig:knn}
\end{figure}

\begin{itemize}
    \item Decision Trees: As illustrated in Figure~\ref{fig:dec}, decision tree is a flowchart-like structure where each internal node represents a decision based on a feature, each branch represents an outcome or decision rule, and each leaf node represents a class label or a final decision~\cite{decisiontreesintro}. The tree is constructed by recursively splitting the data based on the values of input features until a stopping criterion is met, such as maximum depth.

    \begin{figure}[h]
   \centering
   \includegraphics[width=0.5\textwidth]{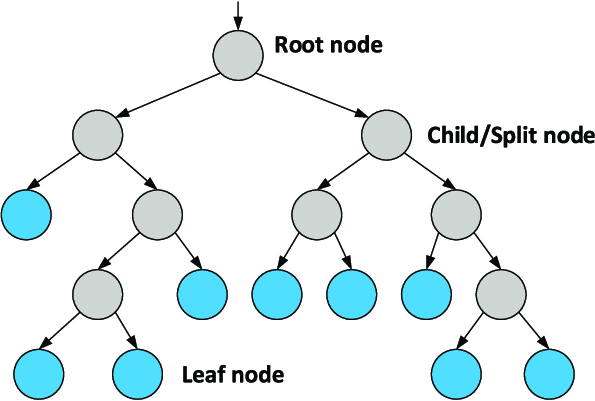}
   \caption{An example illustrating how decision trees are used in classification tasks~\cite{figdecision}. Decision Trees make decisions by recursively partitioning the data set into smaller subsets based on the most discriminative features. The goal is to create branches that lead to homogenous leaves, where each leaf node corresponds to the most probable target outcome. }
   \label{fig:dec}
\end{figure}
    
    \item Naive Bayes is a classification algorithm based on Bayes' theorem, assuming conditional independence of features given the class label~\cite{nbrish2001empirical}. It estimates the likelihood of each feature value for each class in the training dataset. 
    For categorical features, it calculates the probability of occurrence in each class, while for numerical features, it assumes a probability distribution and estimates parameters for each class. By considering prior probabilities and using Bayes' theorem, it calculates posterior probabilities for unlabeled instances. The class label with the highest posterior probability is assigned as the predicted class. 

    \item 
The k-nearest neighbors (KNN) algorithm is a non-parametric classification algorithm based on the principle that similar data points are close to each other in the feature space~\cite{knn}. As illustrated in Figure~\ref{fig:knn}, when a new, unseen instance needs to be classified, the KNN algorithm evaluates the distances between this instance and all other instances in the dataset, identifying the $k$ nearest neighbors. The algorithm then assigns the most frequent label of these nearest neighbors to the new instance.
\end{itemize}

\subsection{Reinforcement Learning}
\label{rl}
\begin{figure}[h]
   \centering
   \includegraphics[width=0.6\textwidth]{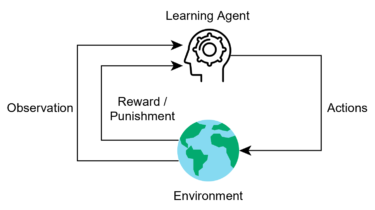}
   \caption{
Reinforcement Learning (RL) involves agents learning to make decisions by taking actions in an environment to maximize cumulative reward. Through trial and error, the agent refines its policy to achieve optimal outcomes. Figure adapted from~\cite{sutton1998reinforcement}.}
   \label{fig:rl}
\end{figure}

Reinforcement Learning (RL) enables an agent to learn and make decisions in an environment through interactions and feedback~\cite{rlwatkins1992qlearning}. 
The agent takes action, receives rewards or punishments, and updates its decision-making strategy accordingly, as shown in Figure~\ref{fig:rl}. 
The RL algorithm's objective is to develop an optimal policy that maximizes cumulative rewards over time. Key components of RL include the agent, environment, state, action, and reward. The agent interacts with the environment by selecting actions based on its current state. The environment provides feedback through rewards or penalties. 
 The agent's decision-making strategy is determined by its policy. The feedback received after taking an action is known as the reward. RL algorithms can be categorized as model-free or model-based. Model-free algorithms directly learn the optimal policy without explicitly modeling the environment, while model-based algorithms learn environment dynamics to plan and make decisions. Notable RL algorithms include Q-Learning, Deep Q-Networks (DQN), Proximal Policy Optimization (PPO), and Advantage Actor-Critic (A2C)~\cite{rlsurvey2, rlsurveykaelbling1996reinforcement}.

\subsection{Deep Neural networks}
\label{dnn}
Deep Neural Networks (DNNs) refer to neural networks with multiple hidden layers between the input and output layers. These hidden layers enable the network to learn hierarchical representations of the input data, allowing for more complex and abstract feature extraction. 
Various categories of DNNs include the following:

\begin{figure}[h]
   \centering
   \includegraphics[width=0.5\textwidth]{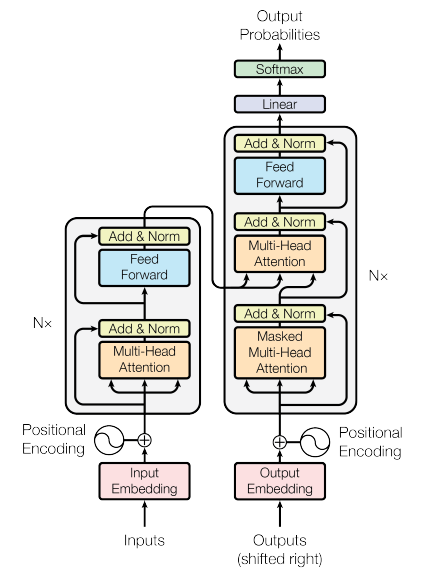}
   \caption{Illustration of the Transformer architecture from \cite{vaswani2017attention}. While this architecture has revolutionized language models, it has been used recently to capture temporal and spatial dependencies in GNSS measurements and improve positioning accuracy.}
   \label{fig:tran}
\end{figure}

\begin{itemize}

\item 
Convolution Neural Networks (CNNs) are widely used in computer vision tasks. They are designed to automatically learn and extract meaningful features from images or other grid-like data through the use of convolutional layers. Convolutional layers apply filters to input data, enabling the network to capture local patterns and spatial dependencies. The pooling layers then downsample the feature maps, reducing their spatial dimensions while retaining important information. Finally, fully connected layers at the end of the network perform classification or regression based on the learned features. CNNs have demonstrated remarkable success in tasks such as image classification, object detection, and image segmentation due to their ability to capture and exploit local patterns and hierarchical representations in visual data.

    \begin{figure}[h]
   \centering
   \includegraphics[width=0.5\textwidth]{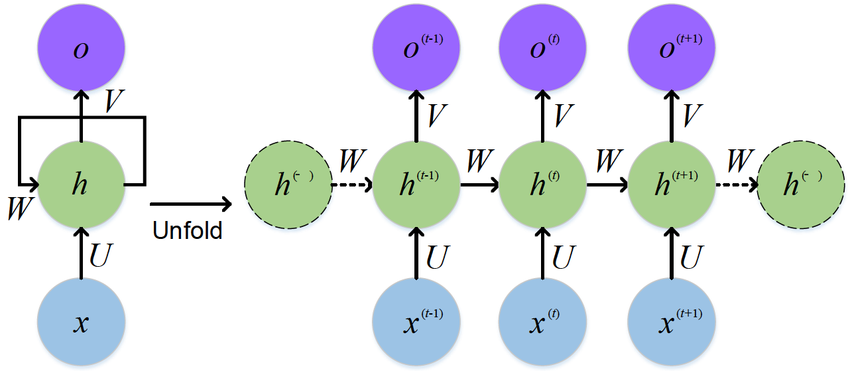}
   \caption{An example RNN architecture from \cite{figrnn}. RNN is a class of neural networks where connections between nodes form a directed graph along a temporal sequence, allowing it to use its internal state or memory to process a sequence of inputs. RNNs process sequential data by maintaining a hidden state that captures information from previous inputs in the sequence. This state is updated at each time step as the network processes the next input, making RNNs ideal for time-series prediction}
   \label{fig:rnn}
\end{figure}

    \item Recurrent Neural Networks (RNNs)~\cite{rnnrumel} are designed to process sequential or time-dependent data. RNNs have feedback connections, allowing information to be fed back into the network at each time step as illustrated in Figure~\ref{fig:rnn}. This recurrent nature enables RNNs to maintain an internal state and capture temporal dependencies. Long Short-Term Memory (LSTM)~\cite{lstmhochreiter1997long} is an RNN architecture specifically designed to model sequential data. Unlike standard feedforward neural networks, which process inputs independently, LSTMs have memory cells that can retain information over time. This memory mechanism makes LSTMs effective in capturing temporal dependencies and long-term patterns in sequential data.

    \item Multilayer Perceptron (MLP)~\cite{mlpbishop1995neural} consists of an input layer, one or more hidden layers, and an output layer. Each neuron in the MLP is connected to neurons in adjacent layers, and these connections have associated weights. MLPs use activation functions to introduce non-linearity into the model, enabling the network to learn complex relationships between the input features and the target variable.
    
    \item Radial Basis Function Neural Network (RBFNN)~\cite{rbfpark1991universal} is a type of neural network that uses radial basis functions as activation functions in its hidden layers. The radial basis functions compute the similarity between the input data and a set of learned prototypes or centers. 
    
    \item Transformer-based deep learning models, as introduced by Vaswani et al.~\cite{vaswani2017attention} and shown in Figure~\ref{fig:tran} use self-attention to capture dependencies among all elements in a sequence concurrently. Through the attention mechanism, these models calculate the significance weights for each element, allowing for effective modeling of relationships between words or tokens. Transformers comprise an encoder and decoder, both consisting of self-attention layers and feed-forward neural networks. The self-attention mechanism employs query, key, and value vectors to compute attention weights and produces outputs that prioritize crucial elements in the sequence.

    \item Graph Neural Networks (GNNs) are a class of deep learning models specifically designed for processing data represented as graphs or networks~\cite{scarselli2009graph, kipf2017semi}. They have gained significant attention in recent years for their effectiveness in various applications, including social network analysis, recommendation systems, and biological network analysis. 
    GNNs handle irregular, graph-structured data by aggregating information from neighboring nodes, enabling them to capture complex relationships and dependencies within the data.
    An example GNN is illustrated in Figure~\ref{fig:gnn}.

    \begin{figure}[h]
   \centering
   \includegraphics[width=0.8\textwidth]{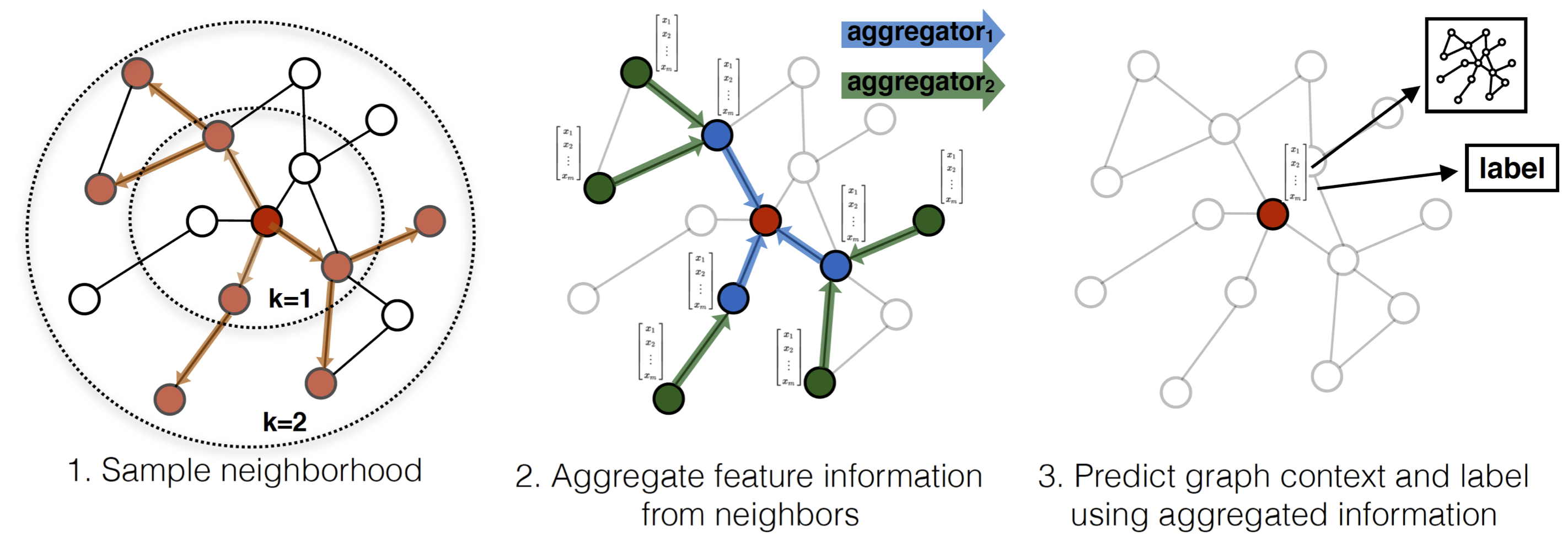}
   \caption{Graph Neural Networks (GNNs) process data on graphs by aggregating information from neighboring nodes. Through iterative updates, they capture complex patterns and relationships inherent in graph structures. An example graph structure (GraphSAGE) is shown here~\cite{graphsage}.}
   \label{fig:gnn}
\end{figure}

\end{itemize}

\subsection{Ensemble methods}
\label{ensemble}
Ensemble methods combine the predictions of multiple individual models to improve overall predictive accuracy and robustness. 
By aggregating the predictions of diverse models, ensemble models can capture different aspects of the data and reduce individual model biases. 
The key categories of ensemble models include the following.

\begin{itemize}
    \item Random Forest combines multiple decision trees to make predictions~\cite{breiman2001random} as shown in Figure~\ref{fig:randomforest}. Each decision tree in the forest is trained on a different subset of the data, and the final prediction is obtained by aggregating the predictions of all trees.

    \begin{figure}[h]
   \centering
   \includegraphics[width=0.6\textwidth]{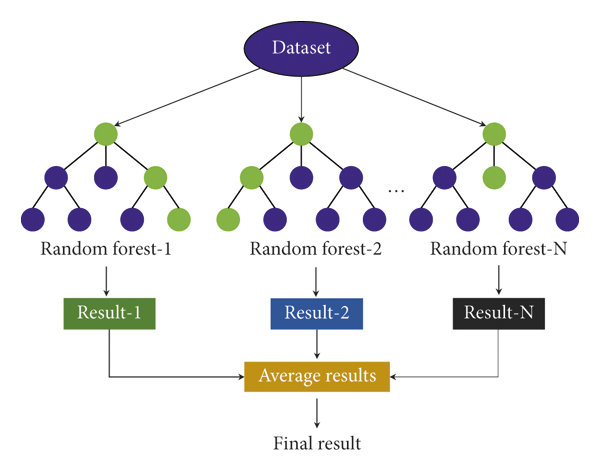}
   \caption{
Random Forest is an ensemble learning method that constructs multiple decision trees during training and outputs the mode of the classes for classification tasks or mean prediction for regression tasks. It introduces randomness by selecting different subsets of features for each tree, improving the model's accuracy and reducing overfitting. Figure adapted from~\cite{figrandomfor}.}
   \label{fig:randomforest}
\end{figure}
    
    \item Gradient Boosting Decision Tree (GBDT) builds trees in a sequential manner, where each new tree is trained to correct the mistakes made by the previous trees~\cite{gradientboostinggreedy}. 
    
    \item LightGBM uses a tree-based learning algorithm similar to GBDT but incorporates several optimizations to speed up training and improve memory efficiency~\cite{lightgbmke2017}. LightGBM supports both classification and regression tasks and has gained popularity for its fast training speed and high performance.
    
    \item Extreme Gradient Boosting (XGBoost) is another gradient-boosting framework that incorporates additional enhancements, such as regularization techniques, to improve model performance~\cite{xgboostchen2016}. XGBoost is known for its flexibility, speed, and its ability to handle various data types.
\end{itemize}

\section{GNSS Signal Analysis and Classification}

\begin{figure}[h]
   \centering
   \includegraphics[width=0.5\textwidth]{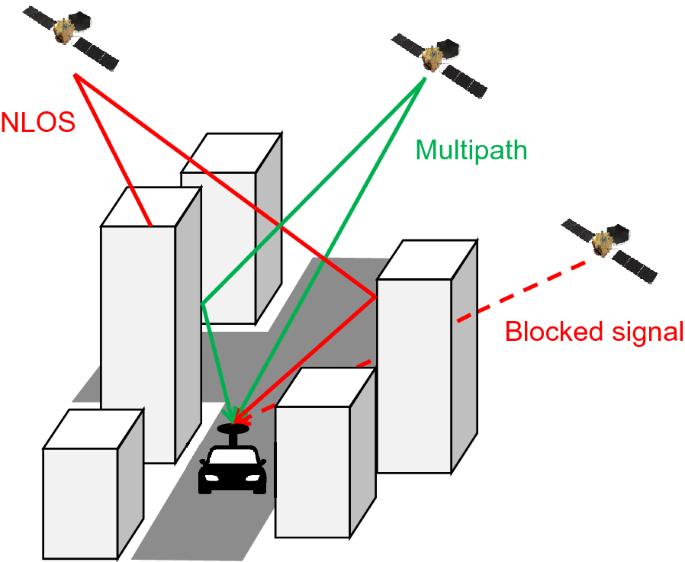}
   \caption{The main error sources in GNSS positioning in urban environments include Non-line-of-sight (NLOS) errors, blocked signals, and MP from reflected signals. By using ML techniques that can classify the signals into these categories, we can improve the accuracy and reliability of positioning. Figure adapted from~\cite{gnss_errors}.}
   \label{fig:gnss_errors}
\end{figure}

In urban environments, GNSS positioning is primarily challenged by MP errors from signal reflections, NLOS errors, and signal blockage due to tall structures.
These error sources are depicted in Figure~\ref{fig:gnss_errors}.
Previous ML techniques have attempted to detect NLOS signals and MP errors as well as classify signals into direct, NLOS, blocked, and MP. 
Several comparative studies have analyzed the efficacy of different ML techniques for these tasks and these studies are explained below.

\textbf{NLOS Detection}: In \cite{socharoentum_machine_2016}, various ML algorithms, including Logistic Regression, SVM, Naïve Bayes, and Decision Tree, were used to detect NLOS signals. Decision Tree and logistic regression models outperformed the other models, achieving an average NLOS prediction correctness rate of 90 \%.
\cite{xu_gnss_2018} demonstrated integrated GNSS shadow matching combined with an intelligent LOS/NLOS classifier based on ML algorithms. Various ML methods were evaluated, achieving classification accuracies between 69.50 \% and 86.47 \% for different urban scenarios. 
Integrating shadow matching with the ML classifier improved positioning accuracy when compared to traditional weighted least squares methods.
For GNSS signal classification and weighting scheme design in built-up areas, \cite{li_machine_2023} proposed an ML-based strategy. The study identified Random forest as the highest-performing classifier for LOS/NLOS classification, achieving a classification accuracy of 93.4 \%. 

\textbf{Time Series Modeling and Prediction}: In \cite{gao_modelling_2022}, ML models, namely GBDT, LSTM, and SVM, were used for the modeling and prediction of GNSS time series. These ML techniques significantly outperformed traditional methods, enhancing the fitting precision by over 30 \%. 

\textbf{MP Detection}: \cite{kim_machine_2022} introduced an ML approach in the context of GPS MP detection leveraging dual antennas. The model, developed using GPS measurements and various algorithms like GBDT, random forest, decision tree, and KNN, achieved classification accuracies between 82 \% to 96 \% for test data from identical training locations. However, the accuracy decreased to 44 \%-77 \% when testing on different locations, with the random forest showing the best classification performance.

\textbf{Monitoring GNSS Satellite Signals for Anomalies}: In \cite{sign_ano}, anomalies were detected in GPS satellite signals using data from globally distributed stations to differentiate between intended and unintended anomalies. Validations involved datasets with known anomalies, testing both supervised and unsupervised algorithms.

We now discuss works that apply either a single technique or combine several ML techniques for enhancing GNSS-based positioning performance.

\subsection{Supervised ML}

\subsubsection{SVM}

Table~\ref{tab:svm_classification}  lists the studies that use SVMs for classification tasks.

\begin{table}[h]
\centering
\caption{GNSS Signal Classification Methods Using SVMs}
\label{tab:svm_classification}
\begin{tabular}{|l|l|c|}
\hline
\textbf{Paper}                           & \textbf{Task}                & \textbf{Accuracy}   \\
\hline
Hsu et al. \cite{hsu}                     & Categorizing pseudorange measurements        & 75\%                \\
\hline
Ozeki et al. \cite{ozeki_gnss_2022}       & NLOS signal detection                        & $>$80\%             \\
\hline
Lee et al. \cite{lee_nonlinear_2022}      & MP prediction model                   & 58.4\% horizontal \\
\hline
Suzuki et al. \cite{suzuki_nlos_2021}     & NLOS MP detection                     & 97.7\%              \\
\hline
Xu et al. \cite{xu_machine_2020}          & GNSS shadow matching in urban environments   & 91.5\%              \\
\hline
\end{tabular}
\end{table}

\textbf{Classifier Design:} In the paper by Hsu et al. \cite{hsu}, a classifier is proposed, trained with SVMs, to categorize GNSS pseudorange measurements into clean, MP, and NLOS categories. Using features extracted from GNSS raw data, the classifier achieves an approximate classification accuracy of 75\%.
In \cite{xu_intelligent_2019}, SVMs are proposed for correlator-level GPS LOS/MP/NLOS signal reception classification, aiming to enhance positioning performance in urban environments. Traditional LOS/MP/NLOS classifiers rely on attributes extracted from basic measurements such as received signal strength and satellite elevation angle. However, their accuracy is limited in urban settings due to complex signal propagation. By extracting LOS/MP/NLOS features at the baseband signal processing stage, the proposed approach achieves improved classification rates, providing valuable insights for enhancing GPS positioning in challenging urban scenarios.

\textbf{MP Prediction}: In Lee et al. \cite{lee_nonlinear_2022}, a MP prediction model based on SVR is designed to improve GNSS performance in deep urban zones. The model factors in the elevation and azimuth angle of each satellite to generate a nonlinear MP map, marking significant improvements of 58.4\% horizontally and 77.7\% vertically in positioning accuracy within a deep urban region in Seoul, Korea.

\textbf{NLOS Detection}: Suzuki et al. \cite{suzuki_nlos_2021} introduced a method to detect NLOS MP by using two supervised learning techniques, SVM and NN. The evaluation shows that NN surpasses SVM and achieves a discrimination accuracy of 97.7\% for NLOS signals.
\cite{losnlosnew}  designed an incremental learning method using an adaptive RBF SVM to detect NLOS signals. The proposed method considers the diversity and complexity of practical factors and shows enhanced performance in harsh canyon cities.
Xu et al. \cite{xu_machine_2020} performed a study on improving the accuracy of GNSS shadow matching in urban environments. They combined a robust estimator with an SVM-based LOS/NLOS classifier. The SVM classifier achieves a classification rate of 91.5\% in urban scenarios.
Ozeki et al. \cite{ozeki_gnss_2022} proposed a method for NLOS signal detection using an SVM classifier trained with unique features derived from receiver-independent exchange format-based information and GNSS pseudorange residual check. By combining the SVM classifier and pseudorange residual check, they achieved more than an 80\% improvement in positioning errors within 10 meters in static tests conducted in dense urban areas.

\textbf{Assessing the Effectiveness of GNSS Features for Signal Classification}: The research in \cite{nlos_new1} is centered around evaluating the efficacy of different GNSS observation features for signal classification using SVMs. The primary metric for evaluation is classification accuracy, and the study is based on an open-source dataset gathered from Hong Kong's urban road segments.
Similarly, \cite{mlnlosnew} performed another study on the efficacy of various features. The authors empirically show the importance of different features in LOS/NLOS signal classification tasks, especially in urban canyon environments.

The literature highlights the success of SVMs in categorizing GNSS pseudorange measurements into clean, MP, and NLOS categories, showcasing their accuracy in signal classification. SVMs have proven to enhance GPS signal reception and processing, particularly in software-defined receivers, outperforming traditional classifiers. They have been successfully utilized for NLOS signal detection and improving positioning accuracy in dense urban areas. By combining SVM classifiers with other techniques, such as pseudorange residual checks and shadow-matching algorithms, significant improvements in positioning accuracy have been demonstrated.

\subsubsection{Decision Trees}

Table~\ref{tab:decision_trees} lists studies that have utilized decision trees for the classification of GNSS signals.

\begin{table}[ht]
\centering
\caption{GNSS Signal Classification Methods Using Decision Trees and GBDT}
\label{tab:decision_trees}
\begin{tabular}{|l|l|c|}
\hline
\textbf{Paper}                             & \textbf{Task}                          & \textbf{Accuracy}   \\
\hline
Guermah et al. \cite{guermah_robust_2018}   & Fusion of left and right antennas & 99\%                \\
\hline
Sun et al. \cite{sun_gradient_2020}         & GPS signal reception classification         & 100\% LOS, 82\% MP, 86\% NLOS \\
\hline
Ye et al. \cite{ye_robust_2020}             & RTK positioning             & 95.64\% NLOS detection\\
\hline
Pan et al. \cite{pan_machine_2023}  & MP mitigation & 24.9 \%- 36.2 \% residual reduction \\
\hline
\end{tabular}
\end{table}

\textbf{Classifier Design:} In the study conducted by Guermah et al.~\cite{guermah_robust_2018}, a signal classifier system is proposed to fuse information from the left and right-polarized antennas using Decision Trees. The classifier achieves an accuracy of 99 \% by utilizing satellite elevation and C/N0 ratio as features, outperforming techniques such as KNN and SVM. Another variant of decision trees, namely the GBDT, is used in Sun et al.'s research~\cite{sun_gradient_2020} for GPS signal reception classification in urban areas, using features such as carrier-to-noise ratio (C/N0), pseudorange residuals, and satellite elevation angle. The GBDT algorithm achieves classification accuracies of 100 \% for LOS signals, 82 \% for MP signals, and 86 \% for NLOS signals, surpassing other algorithms such as decision trees, KNNs, and adaptive network-based fuzzy inference systems.

\textbf{RTK Positioning:} Furthermore, Ye et al.~\cite{ye_robust_2020} designed a robust real-time kinematic (RTK) positioning method that incorporates a decision tree for NLOS signal detection and real-time estimation of double-differenced MP errors. Their method shows remarkable results, achieving an NLOS detection rate of 95.64 \% and enhancing the ambiguity fixing rate by 43 \% in the instantaneous mode. This leads to an approximately 81.77 \% improvement in 3D position accuracy compared to standard RTK methods.

\textbf{MP Mitigation}: In \cite{pan_machine_2023}, the authors proposed a machine learning-based method for mitigating MP in high-precision GNSS data processing. They used XGBoost and formulated MP modeling as a regression task. The XGB-based MP model outperformed conventional methods, achieving substantial residual reduction rates ranging from 24.9 \% to 36.2 \% for various GPS observations. After implementing the XGB-based MP corrections, significant improvements in kinematic positioning precision were observed.

Existing literature shows that decision tree-based classifiers can achieve high accuracy rates and provide robust NLOS signal detection, leading to improved positioning performance. 
However, these classifiers have limitations that should be considered. They are sensitive to feature selection and engineering, requiring careful consideration for optimal performance. Overfitting is a concern, necessitating regularization techniques and model validation. Additionally, decision trees may exhibit instability and lack robustness in the presence of data variations, requiring further exploration of ensemble methods and hybrid approaches. 

\subsection{Unsupervised ML}

The literature on utilizing unsupervised learning techniques to enhance GNSS-based positioning is sparse and limited, however, we discuss a few notable works. 

\textbf{Classifier Design:}\cite{shukla_unsupervised_2022} used an unsupervised ML approach for the classification of NavIC signals affected by MP interference. By leveraging unsupervised learning algorithms, the proposed method classified signals based on unlabeled data, addressing the limitations of supervised learning algorithms that require labeled data. The approach demonstrated promise in detecting and removing MP-affected signals, thereby contributing to more robust positioning applications.

\textbf{MP Detection}: In the study by \cite{savas_multipath_2019}, a MP detection method based on K-means clustering was proposed. The authors applied the K-means algorithm to identify MP signals and evaluated the algorithm's performance in a MP-prone environment. The results indicated that the proposed method exhibited potential for MP detection in GNSS receivers.
Similarly, in \cite{zawislak_gnss_2022}, an unsupervised machine-learning approach for GNSS MP detection was introduced. The method utilizes a CNN within an autoencoder framework combined with k-means clustering. Compared to baseline approaches, the proposed method improved MP detection accuracy and achieved a prediction accuracy of up to 99 \% using unsupervised domain adaptation.

While supervised learning algorithms, such as SVMs and Decision Trees, have been extensively explored and proven effective in GPS signal classification and positioning accuracy improvement, the application of unsupervised learning methods in this domain remains relatively unexplored. Unsupervised learning algorithms, such as clustering or dimensionality reduction techniques, have the potential to discover hidden patterns and structures in GNSS data without the need for labeled training data. By leveraging unsupervised learning, it may be possible to uncover valuable insights and improve positioning performance in novel ways.

\subsection{Deep Learning}

Deep learning approaches are popular for GNSS signal analysis and classification since they can learn directly from raw GNSS signal data, eliminating the need for handcrafted feature engineering. This capability is advantageous in GNSS signal analysis, where the underlying patterns and characteristics may be challenging to define explicitly.
Table~\ref{table:deep} provides an overview of commonly used deep learning approaches for GNSS signal analysis and classification. 

\begin{table}[ht]
\caption{Deep Learning Approaches for GNSS MP Mitigation}
\label{table:deep}
\begin{tabular}{|p{2.5cm}|p{4cm}|p{4cm}|}
\hline
\textbf{Study} & \textbf{Method/Approach} & \textbf{Application/Result}\\
\hline
Suzuki et al. \cite{suzuki_nlos_2021} & SVM and NN-based method for detecting NLOS MP in GNSS. & NN achieved 97.7 \% discrimination accuracy for NLOS signals, outperforming SVM.\\
\hline
Maaref et al. \cite{maaref_leveraging_2021} & DNN-based framework for MP mitigation in GNSS pure L5 receivers. & Significant improvement in positioning accuracy and reduction of pseudorange error standard deviation in heavy MP signal environments.\\
\hline
Orabi et al. \cite{orabi_machine_2020} & NN-based DLL for GPS code phase estimation in high MP environments. & Outperformed conventional techniques in terms of code phase root mean squared error.\\
\hline
Kim et al. \cite{kim_wavelet_2023} & Wavelet transform and NN-based method for GNSS signal quality monitoring and MP detection. & High accuracy in signal quality monitoring and MP detection using real GNSS data.\\
\hline
Li et al. \cite{li_deep_2023} & DNN-based correlation schemes for mitigating MP propagation in GNSS. & Enhanced performance compared to standard correlation schemes in LOS scenarios.\\
\hline
Klimenko et al. \cite{klimenko_evaluation_2021} & Neural network-based MP estimation algorithm for GNSS receivers. & Promising results in compensating for MP errors in GNSS receivers, demonstrating advantages over existing parametric algorithms.\\
\hline
\end{tabular}
\end{table}

\textbf{MP Detection:} In \cite{suzuki_nlos_2021}, the authors proposed a method for detecting NLOS MP using two supervised learning methods, SVM and DNN. 
The evaluation shows that the NN outperforms SVM, achieving a 97.7 \% discrimination accuracy for NLOS signals.
In \cite{maaref_leveraging_2021}, an ML-based framework is developed for mitigating MP in a GNSS pure L5 receiver. They quantified the performance of a pure L5 receiver in static and dynamic heavy MP signal environments and proposed a DNN-based methodology to leverage ML for MP mitigation. The proposed framework significantly improves positioning accuracy and reduces the standard deviation of the pseudorange error.
\cite{orabi_machine_2020} developed a neural network (NN)-based delay-locked loop (DLL) for GPS code phase estimation in MP environments. The proposed NN-based DLL outperforms conventional techniques, including early-minus-late DLL, narrow correlator, and high-resolution correlator, in terms of code phase root mean squared error in high MP environments.
In the paper by \cite{kim_wavelet_2023}, a combination of wavelet transform and neural network is proposed for GNSS signal quality monitoring and MP detection. Signal features, including signal strength and spectral characteristics, are extracted using wavelet transform, while a trained neural network performs classification and MP detection. The proposed method is evaluated using real GNSS data and achieves high accuracy in both signal quality monitoring and MP detection tasks.
In a study by \cite{li_deep_2023}, DNN-based correlation schemes are investigated to mitigate the effects of MP propagation in GNSS. These DNN-based schemes exhibit superior performance compared to standard correlation schemes, particularly in line-of-sight (LOS) scenarios. 
In \cite{li_machine_2023}, the authors also demonstrate that DNN-based correlation schemes outperform standard correlation schemes in line-of-sight scenarios by filtering out more noise and effectively distinguishing MP signals from line-of-sight signals. The proposed DNN-trained models exhibit enhanced performance in time-delay tracking across various realistic scenarios. 
Another research by \cite{klimenko_evaluation_2021} presents a neural network-based MP estimation algorithm for GNSS receivers. The algorithm leverages a 5-point complex correlator implemented in a high-precision GNSS ASIC to mitigate MP errors. Evaluation against existing parametric algorithms demonstrates the algorithm's advantages in accurate MP estimation.

The studies demonstrate that DNN-based methods outperform traditional approaches, such as SVM and conventional correlators, in discriminating NLOS signals and mitigating MP effects. Additionally, the integration of wavelet transform with neural networks shows promise for signal quality monitoring and MP detection.

\subsubsection{CNN}

\begin{figure}[h]
   \centering
   \includegraphics[width=0.7\textwidth]{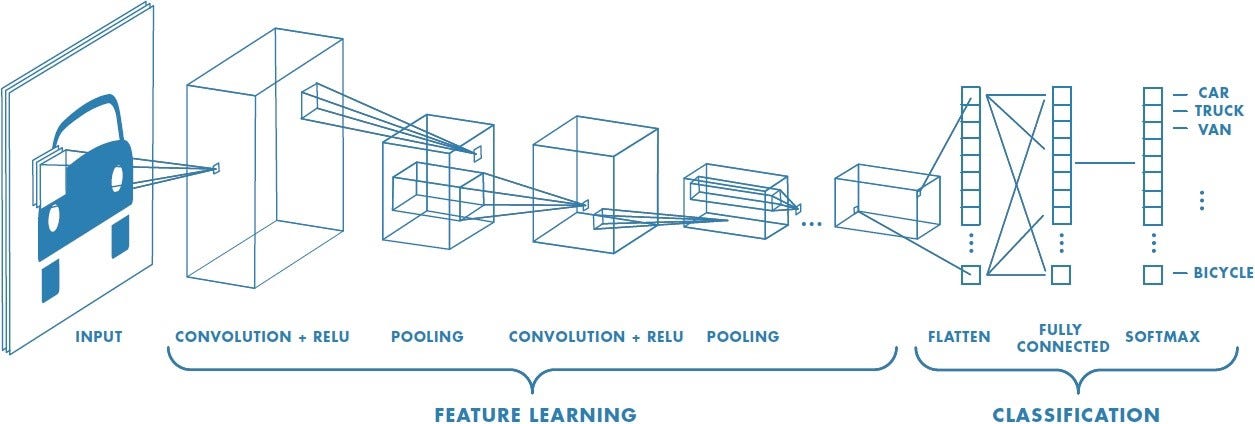}
   \caption{
Convolutional Neural Networks (CNNs) use layered filters to automatically and adaptively learn spatial hierarchies of features from input images. Through pooling and convolution operations, they efficiently recognize and classify visual patterns. Figure adapted from~\cite{visual_cnn}.}
   \label{fig:cnn}
\end{figure}

Several studies have explored the application of CNNs, as illustrated in Figure~\ref{fig:cnn}, for addressing MP and NLOS reception issues and improving positioning accuracy in urban environments.
Table~\ref{table:cnn_MP_signal} provides an overview of these studies.

\textbf{MP Detection:} \cite{munin_gnss_2020} proposed a CNN for mitigating MP in an L5 receiver in urban environments. Experimental results show that the proposed framework significantly improves horizontal positioning accuracy and reduces pseudorange errors. 
Correlator level measurements are used in \cite{munin_convolutional_2020} along with CNNs for MP detection in GNSS receivers. 
The correlator output signal is mapped as a 2D input image, and a CNN is trained to automatically extract relevant features and achieve MP detection. 
In \cite{blais_novel_2022}, the correlation outputs of GNSS signals are mapped into 2D grayscale images, which are fed into a CNN for automatic feature extraction and MP pattern detection. 
The proposed CNN-based algorithm demonstrates superior performance over the benchmark SVM technique, achieving prediction accuracy of over 93 \% even under poor receiving conditions.
The proposed CNN-based algorithm demonstrates superior performance over the benchmark SVM technique, achieving prediction accuracy of over 93 \% even under poor receiving conditions.
CNNs are used for MP detection in both static and kinematic settings in \cite{quan_convolutional_2018}. The proposed method leverages the ability of CNNs to learn and identify the features of MP characteristics from MP-contaminated GPS data. The results demonstrate that the CNN-based method can detect approximately 80 \% of MP errors, leading to improved positioning accuracy when down-weighting the detected MP measurements.
\cite{guillard_using_2023} developed a CNN-based approach to detect GNSS MP using only correlator outputs. The CNN was trained on images representing correlator output values as a function of delay and time. 
The proposed model achieved F scores of 94.7 \% for Galileo E1-B and 91.6 \% for GPS L1 C/A, demonstrating its effectiveness in MP detection.
In \cite{xu_positionnet_2022}, a CNN-based approach for GNSS positioning is proposed to mitigate MP NLOS reception issues. It introduces a new input feature called single-differenced residual map, which effectively mitigates MP/NLOS. The network extracts features from residual maps and generates heat maps to indicate the user's location. PositionNet significantly improves positioning accuracy in dense urban areas, achieving 5-meter-level accuracy for 84 \% of the epochs.
In \cite{suzuki_nlos_2020} a novel NLOS MP detection technique is presented using CNNs to improve positioning accuracy in urban environments. The CNN-based NLOS discriminator achieved approximately 98 \% correct discrimination of NLOS MP signals, outperforming a simple neural network. By applying the NLOS probability output of the CNN to positioning calculations, the proposed method improved positioning accuracy from 34.1 to 1.6 meters.

\textbf{Signal Classification:} \cite{elango_disruptive_2022} proposed a robust deep-learning-based technique for detecting and classifying disruptive GNSS signals, including jammers, spoofing, and MP signals. The approach utilized transfer learning with pre-trained CNNs such as AlexNet, GoogleNet, ResNet-18, VGG-16, and MobileNet-V2. The MobileNet-V2 model achieved an accuracy of 99.8 \% in classifying different types of disruptive signals.
In \cite{jiang_convolutional_2022}, a CNN is proposed that utilizes correlator-level measurements. 
They employ vector tracking to generate correlator-level measurements and the CNN automatically extracts features and identifies the signal reception type. 
The proposed CNN outperforms other methods such as KNN and SVMs in terms of classification accuracy.
In \cite{suzuki_nlos_2021}, the authors use GNSS signal correlation output as input for supervised learning methods, specifically SVMs and DNNs, to classify NLOS signals. The evaluation shows that the DNN outperforms SVM, achieving 97.7 \% correct discrimination of NLOS signals.
For smartphone-based positioning, \cite{liu_nlos_2022} designed a method to detect and correct NLOS signals utilizing a CNN that achieves enhanced positioning accuracy in urban environments.

\begin{table}[ht]
\caption{CNN Methods for GNSS MP Mitigation and Signal Analysis}
\label{table:cnn_MP_signal}
\begin{tabular}{|p{2.5cm}|p{4cm}|p{4cm}|}
\hline
\textbf{Study} & \textbf{Approach} & \textbf{Result}\\
\hline
Munin et al. \cite{munin_gnss_2020} & CNN-based framework for mitigating MP in a pure L5 GNSS receiver  & Significant improvement in horizontal positioning accuracy \\
\hline
Elango et al. \cite{elango_disruptive_2022} & Transfer learning with pre-trained CNN models for detecting and classifying disruptive GNSS signals. & 99.8 \% accuracy in classifying disruptive signals\\
\hline
Jiang et al. \cite{jiang_convolutional_2022} & CNN-based method for signal classification using correlator-level measurements. & CNN outperformed KNN and SVM in terms of classification accuracy.\\
\hline
Liu et al. \cite{liu_nlos_2022} & CNN-based method for NLOS signal detection and correction in smartphone-based positioning. & Enhanced positioning accuracy and stability in urban environments.\\
\hline
Li et al. \cite{li_machine_2023} & DNN-based correlation for mitigating MP propagation. & Improved performance in time-delay tracking and MP signal classification\\
\hline
Blais et al. \cite{blais_novel_2022} & CNN-based method for MP prediction using correlation outputs of GNSS signals. & Superior performance over SVM in MP prediction even under poor receiving conditions.\\
\hline
Guillard et al. \cite{guillard_using_2023} & CNN-based approach using correlator outputs for GNSS MP detection. & High accuracy in classifying signals as line-of-sight (LOS) or MP, outperforming traditional ML classification models.\\
\hline
Quan et al. \cite{quan_convolutional_2018} & CNN-based method for MP detection in static and kinematic GNSS settings. & 80 \% detection of MP errors\\
\hline
Suzuki et al. \cite{suzuki_nlos_2021} & NN-based method for detecting NLOS MP in GNSS signals. & NN achieved 97.7 \% correct discrimination of NLOS signals, outperforming SVM.\\
\hline
Liu et al. \cite{liu_nlos_2022} & CNN-based approach using single-differenced residual map & 5 m accuracy for 84 \% epochs.\\
\hline
\end{tabular}
\end{table}

Based on the discussed papers, several key insights emerge regarding the effectiveness of CNNs in GNSS signal analysis and classification. Firstly, CNNs show promise in mitigating MP effects, resulting in notable enhancements in horizontal positioning accuracy and reduced pseudorange errors. Secondly, CNNs exhibit strong capabilities in detecting and classifying MP signals, achieving high accuracy rates. Thirdly, CNN-based regression models outperform traditional methods in GNSS MP estimation, enabling uncertainty modeling and maintaining estimation performance even with lower input image resolution. Fourthly, CNNs excel in signal classification tasks utilizing correlator-level measurements, surpassing alternative approaches.
Lastly, CNNs contribute significantly to NLOS signal detection and correction, leading to improved positioning accuracy and stability in urban environments.

\subsubsection{RNN}
We summarize key papers in Table~\ref{tab:rnn_classification} that utilize RNNs for GNSS signal analysis and classification.

\begin{table}[h]
\centering
\caption{GNSS NLOS/LOS Classification Methods Using RNN}
\label{tab:rnn_classification}
\begin{tabular}{|l|l|c|}
\hline
\textbf{Paper}                                & \textbf{Classification Task}                          & \textbf{Accuracy} \\
\hline
Su et al.\cite{su_non-line--sight_2022}                 & NLOS/LOS classification    & 91\%              \\
\hline
Cho et al.\cite{cho_enhancing_2019}                      & NLOS/LOS classification in urban environments         & 90\%              \\
\hline
Lyu et al.\cite{lyu_new_2020}                            & LOS/NLOS signal classification     & 95.97\%           \\
\hline
Liu et al.\cite{grunew}                            & Context classification    & 99.41\%           \\
\hline
\end{tabular}
\end{table}

\textbf{NLOS classifier:} In \cite{su_non-line--sight_2022}, the authors proposed an NLOS/LOS classification model based on RNNs to classify satellite signals received in urban canyon environments. 
The model achieves an accuracy of 91 \% in classification and demonstrates improved three-dimensional positioning accuracy and stability in the BDS/GPS fusion system. 
The proposed method outperforms traditional ML classification models like SVMs. 
\cite{cho_enhancing_2019} proposed an RNN-based NLOS classifier that discriminates between LOS and  NLOS satellites in urban environments. The classifier achieved about 90 \% accuracy in NLOS classification and showed a 20 \% improvement in discrimination performance compared to the conventional SVM-based NLOS classifier. The proposed technique was also applied to pedestrian road crossing detection and demonstrated a positioning accuracy of about 45 \% better than that of conventional techniques.
\cite{lyu_new_2020} proposed a hybrid RNN and fully connected network approach to distinguish between LOS and NLOS signals in GNSS positioning. The method considered inter-epoch information and time series data features to enhance classification accuracy. The proposed classifier achieved an overall testing accuracy improvement from 93.00 \% to 95.97 \% for Rinex-level observations.

\textbf{Context Recognition:} A Gated Recurrent Unit (GRU) for real-time processing is proposed in \cite{grunew} to categorize fine-grained contexts based on the characteristics of different environments and their corresponding integrated navigation method.
The proposed method enhances context recognition using a new feature called the C/N0-weighted azimuth distribution factor and achieves a recognition accuracy of 99.41 \% on a real-world urban driving dataset.
Xia et al. \cite{xia_recurrent_2020} proposed a scenario recognition method based on RNN and LSTM models, utilizing smartphone GNSS measurements. Their analysis focuses on the impact of multi-constellation satellite signals on scenario recognition performance. The results indicate that the accuracy of scenario recognition improves with an increased number of constellations received by smartphones. The proposed algorithm achieves an impressive recognition accuracy of 98.65 \% and effectively handles scenario transitions with a maximum delay of only 3 seconds.

The papers discussed highlight the effectiveness of RNNs for NLOS/LOS classification and GNSS positioning in urban environments. These RNN-based models achieve high classification accuracy, leading to improved positioning accuracy compared to models like SVMs which do not consider inter-epoch information and time series data features.

\subsection{Hybrid Approach}

In \cite{xu_machine_2020}, robust estimation and ML techniques are combined for LOS/NLOS classification and improving shadow matching in urban GNSS positioning. 
The proposed approach utilizes a robust estimator for initial positioning and an SVM for satellite visibility classification. The classification rate of the SVM reaches 91.5 \% in urban scenarios, contributing to improved shadow-matching accuracy.

Hybrid methods have also been recently proposed in other domains, for example, in the context of Kalman filtering and particle filtering which could be directly applied to improving GNSS positioning accuracy.
KalmanNet~\cite{kalmannet} introduces a novel approach to real-time state estimation for dynamical systems with non-linear dynamics or partial information, merging the classic Kalman filter's structure with a recurrent neural network to learn from data. This hybrid model enhances the traditional filter's capabilities, allowing it to adapt to complex dynamics and outperform conventional filtering methods, regardless of the accuracy of the domain knowledge.
In \cite{learnkfun1}, the authors introduce an unsupervised learning adaptation for KalmanNet, a deep neural network system inspired by the Kalman filter, eliminating the need for ground-truth states by using its hybrid architecture to predict observations and compute loss. It demonstrates that unsupervised KalmanNet can match the performance of its supervised counterpart and adapt to changing state space models without new data, showcasing flexibility and efficiency in dynamic environments.
Similarly, in \cite{learnkfun2}, a novel approach called DANSE is introduced for non-linear state estimation, offering a model-free method to compute the posterior state in a Bayesian framework with linear measurements. By employing recurrent neural networks to capture non-linear dynamics and utilizing a combination of maximum likelihood and gradient descent for unsupervised training, DANSE operates effectively without process model knowledge. Its performance is demonstrated to be competitive with both classic model-based estimators.

The authors in \cite{learnpf1} introduced a particle filter RNN (PF-RNN) architecture that combines an advanced RNN architecture with uncertainty modeling by maintaining a distribution of latent states represented as particles. This approach contrasts with traditional RNNs' single deterministic latent vector. PF-RNNs leverage a differentiable particle filter mechanism for updating the latent state distribution in line with Bayes' rule, enhancing the model's adaptability to variable and multi-modal data. 
In \cite{learnpf2}, a particle filter network is designed that integrates a system model and particle filter algorithm into a unified, fully differentiable neural network.
Although it has been only applied to visual localization tasks, it has demonstrated superior performance and generalization over traditional and alternative learning-based approaches, adapting effectively to various and unseen sensor inputs.
Hybrid particle filters were also explored in~\cite{learnpf3} wherein neural networks were integrated with particle filters for scalable real-world applications, focusing on the challenge of optimizing dynamic and measurement models without access to expensive or unavailable true states.
By utilizing a differentiable implementation of particle filters and an end-to-end learning objective based on maximizing a pseudo-likelihood function, the approach improves state estimation accuracy even when true states are largely unknown. The effectiveness of this method is evaluated through state estimation tasks in robotics, using both simulated and real-world datasets.

\begin{figure}[ht]
\centering
\begin{tikzpicture}
\begin{axis}[
    ybar,
    bar width=0.5cm,
    ylabel={Number of Papers},
    ymin=0,
    xtick=data,
    xticklabels={
       SVM,
       Decision Trees,
        Unsupervised Learning,
        Deep Learning,
        CNN,
       RNN,
        Hybrid
    },
    x tick label style={rotate=45, anchor=east},
    ]
\addplot coordinates {
    (1,9)
    (2,4)
    (3,3)
    (4,7)
    (5,12)
    (6,5)
    (7,3)
};
\end{axis}
\end{tikzpicture}
\caption{Distribution of Papers on ML Methods for GNSS Signal Analysis and Classification}
\label{fig:gnss_ai_methods}
\end{figure}
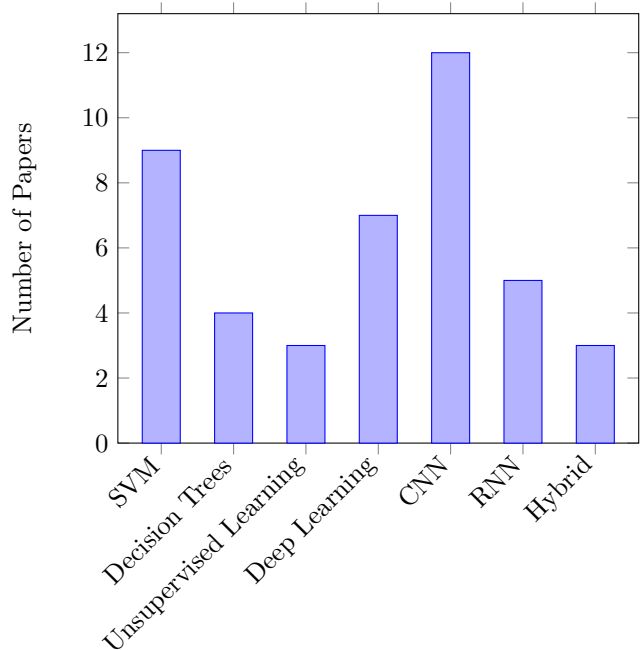

\subsection{Other approaches}
In \cite{satvis1}, a novel approach is proposed for predicting and eliminating MP errors, particularly in urban areas with complex signal reflections. 
The proposed method utilizes a graph transformer neural network (GTNN) to effectively learn environment representations from irregular GNSS measurements. 
Experimental results on real-world GNSS data show that the GTNN achieves over 96 \% accuracy in satellite visibility prediction and outperforms existing MP prediction methods in terms of generalization performance. 
In \cite{nlosours}, a novel method using Neural City Maps, built on Neural Radiance Fields, is proposed to represent urban geometry more accurately. The study evaluates different prediction methods for NLOS effects using Neural City Maps and demonstrates their effectiveness in improving localization accuracy in challenging urban environments.

\section{Environmental Context and Scenario Recognition}

We discuss ML approaches that are used for scenario recognition and environmental context detection.

Supervised machine learning is used in the study by Baldini et al. \cite{baldini_experimental_2021} to train a classifier for propagation scenario identification. The researchers extracted various features from GNSS pseudorange measurements and demonstrated the classifier's accurate identification of propagation scenarios affected by MP. The adoption of an overlapping window approach further enhances identification accuracy. In the work by İşik et al. \cite{isik_machine_2021}, a machine learning-based performance prediction algorithm for GNSS in urban air mobility applications is presented. The algorithm considers environmental parameters and evaluates the prediction performance of three algorithms: KNN, SVR, and Random Forest. The researchers analyze the performance prediction results and the importance of parameters across different urban environments using synthetic data generated by a GNSS simulator.

\cite{scene1} explores the utilization of GNSS signals alongside ML algorithms to characterize the operational environment for Unmanned Aerial Vehicles and Unmanned Ground Vehicles by extracting features relevant to situational awareness in urban and harsh conditions. It presents case studies demonstrating how digital signal processing techniques combined with unsupervised and supervised ML algorithms (like K-means and SVMs) can analyze GNSS observables to identify propagation scenarios affected by MP, interference, and atmospheric conditions. The study in~\cite{scene2} focuses on enhancing GNSS accuracy for train localization by identifying environmental characteristics across scenarios like tunnels, open areas, and urban canyons. Utilizing NMEA-0183 protocol data from GNSS receivers, such as PRN codes, azimuth, elevation, and SNR, the research creates heatmap states of these scenarios through satellite observations, interpolation, and position transformation. By training a Vision Transformer model on these heatmap datasets, the study successfully recognizes varying environmental scenarios, achieving an overall model accuracy of 88.7 \% on the validation set.

In~\cite{scene3}, the authors propose to improve high-precision GNSS positioning for intelligent transportation systems through a context-aware model, addressing the challenges posed by complex environmental contexts and feature variability. The study evaluates eight models, including various neural networks and SVM, for their ability to recognize context. The LSTM model outperforms others, achieving high accuracy and mean average precision in distinct and continuous context areas. The study in~\cite{scene4} introduces a novel signal-based environment recognition algorithm designed for vehicular positioning in urban settings, capable of distinguishing between six distinct environmental conditions. By constructing a signal feature vector that encapsulates signal attenuation, blockage, and MP effects, the algorithm leverages the SVM for scene classification. To improve accuracy, a temporal filtering technique is integrated, allowing the model to adapt and function in real-time for the receiver. Demonstrating the algorithm's broad applicability, datasets for both training and testing were gathered from various cities, achieving an overall recognition accuracy of 89.3 \% across diverse environments. In~\cite{scene5}, the authors improve scenario recognition for mobile applications by classifying environments into four categories and using a Hidden Markov Model and an RNN. The RNN method effectively handles scenario transitions and environmental changes, achieving an overall accuracy of 98.65 \% and a transition recognition accuracy of 90.94 \%, with minimal transition delay. \cite{scene6} introduces a deep-learning method for scenario recognition using smartphone GNSS measurements, categorizing environments into four types: deep indoors, shallow indoors, semi-outdoors, and open outdoors. Leveraging Voronoi tessellations for spatial structuring and employing CNNs and ConvLSTM networks for feature extraction and sequence processing, the technique achieves accuracies of 98.82 \% with CNNs and 99.92 \% with ConvLSTMs. This approach, relying solely on GNSS measurements without additional sensors, demonstrates both efficiency and suitability for real-time applications, with minimal computational latency.

\section{Anomaly Detection and Quality Assessment}
The characterization and assessment of signal quality in multi-GNSS systems are crucial for enhancing the performance of GNSS-based positioning. 
In \cite{quan_new_2017}, the focus is on evaluating measurement signal quality and developing an ML-based MP detection model for multiple GNSS systems, including GPS, GLONASS, Galileo, BDS, and QZSS. The proposed model achieves high accuracy rates using simulated and real GNSS data.
Additionally, \cite{xia_anomaly_2020} combined clustering-based anomaly detection with supervised classification to improve positioning accuracy significantly in different directions. These studies highlight the use of ML-based approaches for evaluating signal quality, detecting anomalies, and enhancing GNSS positioning performance.

M. Kiani introduces a machine learning algorithm tailored for GNSS position time series prediction, demonstrating superior accuracy in outlier and anomaly detection as well as earthquake prediction capabilities by analyzing over three thousand GNSS station time series globally~\cite{new_ano1}. This method outperforms seventeen other algorithms and offers practical applications in detecting time series outliers and earthquake forecasting, exemplified by the Tohoku 2011 case study.
In~\cite{new_ano2}, the authors explore enhancing GNSS signal anomaly detection for navigation systems using time-delayed neural networks (TDNN), proposing a TDNN-based integrity monitoring system that significantly outperforms standard receiver autonomous integrity monitoring (RAIM) methods in speed and reliability. 
An innovative approach for automatic anomaly detection is proposed in~\cite{new_ano3} for monitoring GNSS reference stations. The authors use predictive modeling and statistical rules to identify anomalous signals, demonstrating the method's effectiveness on historical data.
ML algorithms such as random forest~\cite{new_ano4} and SVMs~\cite{new_ano6} are used to detect GPS and Galileo satellite oscillator anomalies, respectively, with high accuracy, outperforming other algorithms and demonstrating global applicability for satellite anomaly monitoring.

Unsupervised ML methods are used for autonomous GNSS data anomaly detection in~\cite{new_ano5}, focusing on volcanic activity monitoring. 
Unsupervised methods are also used for spatial outlier detection in GNSS velocity fields using a robust Mahalanobis-distance-based classification method~\cite{new_ano7}. Their approach, validated on synthetic and real datasets, yields high classification accuracy, enhancing GNSS data reliability without requiring pre-defined labels.

\section{GNSS Integration with Other Sensors}

The integration of ML methods in GNSS-based positioning systems, particularly in combination with other sensors like inertial measurement units (IMUs), has opened up new possibilities for improving accuracy and addressing challenges in various environmental contexts. We discuss some notable works below.

\textbf{NLOS Detection: } In their paper, Wang et al. \cite{wang_multipathnlos_2022} introduced a method that utilizes the K-means clustering algorithm to detect MP and NLOS signals in urban areas for GNSS/INS integrated positioning. The method incorporates various feature parameters derived from GNSS raw observations and demonstrates significant improvements in positioning accuracy. The offline dataset exhibits a remarkable improvement of 16 \% and 85 \% in the horizontal and vertical directions, respectively, while the online dataset showcases improvements of 21 \% and 41 \% in these two directions.
Smolyakov et al.

\textbf{MP Prediction: }\cite{smolyakov_resilient_2020} proposed a two-part architecture for GNSS MP prediction and detection in IMU/GNSS integration for urban navigation. It employs signal quality monitoring techniques to identify and exclude MP-contaminated GNSS signals. The architecture dynamically adjusts the integration Kalman filter based on a crowdsourced GNSS MP environment map, which is extended to unsurveyed areas using a random forest ML model. Evaluation in an automotive scenario shows a significant accuracy improvement compared to a conventional Kalman filter (13-17 \%).

\textbf{Positioning Improvement:} Han \cite{hanrl} proposed a reinforcement learning-based approach to optimize the process noise covariance matrix of a GNSS/IMU integration Kalman filter. Experimental results show improved navigation performance by effectively utilizing the learned process noise covariance matrix. Additionally, Shin et al. \cite{shin_implementation_2018} designed an Actor-Critic (A2C) Reinforcement Learning algorithm that achieves higher scores compared to the baseline.
Gao et al. \cite{gao_rl-akf_2020} presented the RL-AKF (adaptive Kalman filter) navigation algorithm, which adaptively estimates the process noise covariance matrix using a reinforcement learning approach. The RL-AKF demonstrates an average positioning error of 0.6517 m within a 10 s GNSS outage for the GNSS/INS integrated navigation system. For the GNSS/INS/Odometer (ODO) and GNSS/INS/Non-Holonomic Constraint (NHC) integrated navigation systems, the RL-AKF achieves positioning errors of 14.9426 m and 15.3380 m, respectively, within a 300 s GNSS outage.
In their study, Li et al. \cite{li_gnssins_2022} enhanced the GNSS/INS integration methodology for vehicle navigation by using the LightGBM regression model. This model predicts vehicle position changes during GNSS outages based on INS data. The proposed methodology demonstrates reduced errors in predicting vehicle positions during GNSS outages compared to the existing methodology based on Random Forest. The integration of artificial intelligence improves the accuracy of GNSS/INS integrated navigation systems in situations where GNSS signals are unavailable or during GNSS outages.
Chiou et al. \cite{chiou_performance_2019} developed an ML model to enhance the utilization of GNSS positions in a loosely coupled GNSS/IMU system. The proposed model combines rule-based methods with machine learning techniques to classify the quality of GNSS position outputs. The results show that the model achieves a true positive rate of 90 \% in identifying bad GNSS position outputs.
In \cite{realkf}, the authors integrated GNSS and INS sensors using deep learning techniques. 
They combine DNN, LSTM, and CNN to optimize Kalman filter gain and improve navigation accuracy for land vehicles.

The papers in this section present valuable contributions to the field of GNSS integration with other sensors for navigation in urban environments. These contributions include the application of ML techniques, such as clustering algorithms and reinforcement learning, to enhance positioning accuracy. 
The development of dynamic sensor integration models based on environmental maps and MP detection techniques offers improved performance.

\section {Prediction and Forecasting}
In the domains of geodesy and GNSS analysis, ML methods have proven to be instrumental in prediction and forecasting tasks.

\textbf{Time Series Prediction: }In their study, Shahvandi et al. \cite{shahvandi_modified_2021} used deep transformers to predict time series in the field of geodesy. They make modifications to the original network architecture and optimization procedure, resulting in a remarkable improvement of 21.5 \% in prediction accuracy compared to traditional statistical methods. Furthermore, their approach outperforms other machine learning algorithms by at least 2.7 \%. The method exhibits the potential to achieve millimeter accuracy in time series prediction.
Loli Piccolomini \cite{loli_piccolomini_recurrent_2019} introduced a network architecture based on LSTMs for denoising and prediction tasks in GNSS time series analysis. 
Despite being a shallow network, it reduces scattering from real GNSS time series, removing nearly 50 \% of the noise. Additionally, the architecture achieves coordinate prediction with a mean squared error of 1.1 millimeters. The approach is evaluated using both synthetic and real GNSS time series data.
In their research, Ji et al. \cite{ji_signal_2020} presented a weighted wavelet analysis-based signal extraction method for GNSS position time series. This method successfully extracts signals from daily position time series data by considering noise characteristics and variations in signal strength. The application of weighted wavelet analysis enhances the accuracy of signal extraction, particularly in the presence of noise and disturbances.

\textbf{Satellite Visibility Prediction: }Zhang et al. \cite{zhang_prediction_2021} proposed a deep learning network architecture that combines fully connected neural networks (FCNNs) and LSTM networks to predict GNSS satellite visibility and pseudorange error based on GNSS measurement-level data. The proposed networks achieve an accuracy of 80.1 \% in satellite visibility prediction and an average difference of 4.9 meters in pseudorange error prediction. The LSTM layer effectively captures representations of the environment, leading to improved prediction performance.

\section {Position Error Modeling/Accuracy Enhancement}
A significant body of work focuses on utilizing ML techniques to model GNSS errors in the position domain and enhance positioning accuracy directly. These works form the majority of research efforts in this field and are summarized in Table~\ref{table:ai_gnss_1} and Table~\ref{table:ai_gnss_2}.
Figure~\ref{fig:gnss_ai_methods} shows the distribution of different ML methods that have been used to improve GNSS positioning accuracy.
These various approaches highlight the effectiveness of ML techniques in enhancing GNSS positioning accuracy and addressing specific challenges in different domains and environments.

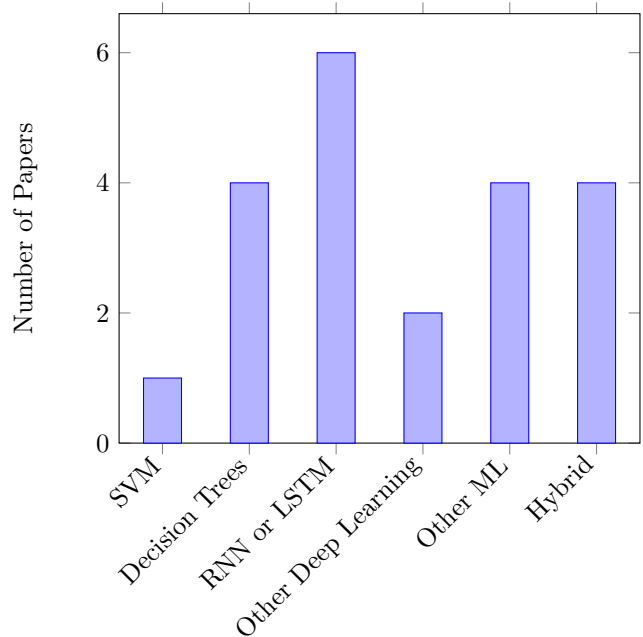
\begin{figure}[ht]
\centering
\begin{tikzpicture}
\begin{axis}[
    ybar,
    bar width=0.5cm,
    ylabel={Number of Papers},
    ymin=0,
    xtick=data,
    xticklabels={
       SVM,
       Decision Trees,
       RNN or LSTM,
        Other Deep Learning,
       Other ML,
        Hybrid
    },
    x tick label style={rotate=45, anchor=east},
    ]
\addplot coordinates {
    (1,1)
    (2,4)
    (3,6)
    (4,2)
    (5,4)
    (6,4)
};
\end{axis}
\end{tikzpicture}
\caption{Distribution of Papers that use ML methods for Position Error Modeling and Accuracy Enhancement}
\label{fig:gnss_ai_methods}
\end{figure}

\textbf{Improving PPP and RTK}: Qafisheh et al. \cite{qafisheh_support_nodate} utilized SVMs to reduce latency in real-time Precise Point Positioning (PPP), leading to improved clock corrections. Menzori and Teunissen \cite{menzori_evaluation_2018} adopt Decision Trees for classifying PPP/GNSS coordinates based on precision. Additionally, Yun et al. \cite{yun_practical_2022} proposed leveraging dual-frequency GNSS measurements, and Mendonca et al. \cite{mendonca_rtk_2022} introduced a Genetic Algorithm (GA)-based machine learning classifier to improve RTK positioning.
Lacambre et al. \cite{lacambre_optimizing_2022} designed machine learning and outlier detection methods to optimize RTK positioning, achieving a substantial boost in real-world positioning performance.

\textbf{Improving Accuracy in MP Environments}: Ziedan et al. \cite{ziedan_optimized_2021} proposed two novel ML-based algorithms that make use of maps and neural networks to accurately estimate positions in MP environments. The paper by \cite{cognss} introduced a Random Forest-based technique to rectify errors in differential corrections for Cooperative Differential GNSS in urban landscapes. This approach aims to predict and amend biases stemming from positioning errors of cooperative vehicles, enhancing overall accuracy. Urban road tests demonstrate a reduction in RMSE from 1.557 to 1.279 m using reference stations. Sun et al. \cite{sun_improving_2021} addressed the challenges of achieving accurate positioning in urban terrains. They introduced a GBDT-based method that predicts pseudorange errors and corrects positioning inaccuracies caused by MP and NLOS signals, achieving a marked 70 \% improvement in 3D positioning accuracy compared to traditional methodologies.

\begin{table}[h]
\caption{ML Methods for Improving GNSS Positioning Accuracy}
\label{table:ai_gnss_1}
\begin{tabular}{|p{2.5cm}|p{4cm}|p{4cm}|}
\hline
\textbf{Study} & \textbf{Method/Approach} & \textbf{Application/Result}\\
\hline
Qafisheh et al. \cite{qafisheh_support_nodate} & SVM-based solution for latency reduction in PPP & Reduced standard deviation and range of clock corrections by approximately 30 \% and 20 \%, respectively\\
\hline
Ziedan et al. \cite{ziedan_optimized_2021} & ML-based algorithms improve position estimation accuracy in MP environments. & Accuracy enhancements of up to 96 \% compared to traditional methods\\
\hline
Mendonca et al. \cite{mendonca_improving_2022} & ML algorithms (decision tree, neural network, etc.) to enhance integrity measurements in GNSS positioning. & Neural network model increased information metric by 20-fold compared to EKF.\\
\hline
Menzori and Teunissen \cite{menzori_evaluation_2018} & Decision Tree classification of the accuracy of PPP/GNSS coordinates  & Prediction of accuracy from the precision with a large dataset of known coordinates.\\
\hline
Kim et al. \cite{kim_deep_2019} & Multilayer RNN with LSTM algorithm  & Improved position accuracy by 40 \% compared to GNSS-only navigation.\\
\hline
Yang et al. \cite{yang_build_2019} & Real-time LSTM RNN for predicting GPS positioning errors & Prediction accuracy within 1-3 \% of ground truth values.\\
\hline
Gupta et al. \cite{gupta_designing_2022} & Hybrid learning-based approach combining traditional positioning models with DNNs & Low positioning errors with reduced memory requirements.\\
\hline
Kanhere et al. \cite{kanhere_improving_2022} & DNN-based corrections using set transformer& Improved accuracy over WLS\\
\hline
Dai et al. \cite{dai_2nd_2022} & Global optimization method incorporating various constraints for smartphone positioning  & Second place in Google Smartphone Decimeter Challenge 2022.\\
\hline
Liu et al. \cite{liu_vehicular_2019} & LSTM-based prediction method  & Improved prediction accuracy by 16 \% .\\
\hline
Thomas et al. \cite{thomas_position_2021} & ML-based post-processing techniques for low-cost GPS receivers  & Improved position accuracy \\
\hline
Zhou et al. \cite{zhou_least-squares_2017} & LSSVM-KF algorithm for by estimating dynamic modeling bias. & Reliable and accurate GNSS navigation solutions\\
\hline
Gao et al. \cite{gao_method_2022} & Decision tree model for estimating vehicle positioning accuracy & Achieved probability of accurate positioning estimation of more than 95 \% \\
\hline
Wei et al. \cite{wei_satellite_2016} & GRNN-based satellite selection algorithm for optimizing visible satellites & Improved robustness, accuracy, and real-time performance \\
\hline
\end{tabular}
\end{table}

\textbf{Improving Accuracy for Smartphone Positioning}:
Google has also shown active involvement in refining urban GNSS accuracy, focusing on mitigating inaccuracies experienced by Android devices in dense urban settings, a vital endeavor for widely-used location apps \cite{noauthor_google_2020}. 
Hybrid methods that combine traditional models with deep neural networks are introduced by Gupta et al. \cite{gupta_designing_2022} and Kanhere et al. \cite{kanhere_improving_2022} with the aim to enhance data efficiency and positional accuracy for smartphones. Additionally, Dai et al. \cite{dai_2nd_2022} presented a global optimization strategy for smartphone GNSS positioning. 
Lastly, in \cite{meold}, \cite{menew}, and \cite{gnnrnn}, the authors proposed advanced GNSS solutions using Graph Convolution Neural Networks and a combination of RL and GNN, demonstrating significant improvements in smartphone positioning accuracy in various environments.

\begin{table}[h]
\caption{ML Methods for Improving GNSS Positioning Accuracy (Continued)}
\label{table:ai_gnss_2}
\begin{tabular}{|p{2.5cm}|p{4cm}|p{4cm}|}
\hline
\textbf{Study} & \textbf{Method/Approach} & \textbf{Application/Result}\\
\hline
Ragheb et al. \cite{ragheb_gnss_2021} & SlipNet LSTM neural network model for cycle slip detection. & High-performance results with 99.7 \% detection and localization accuracy\\
\hline
Neri et al. \cite{neri_machine_2020} & ML architecture for local hazard detection and RAIM in the rail domain. & Optimizing ML architecture for enhancing the accuracy and reliability of train positioning systems.\\
\hline
He et al. \cite{he_research_2023} & LSTM neural network model for BDS-3 satellite clock bias prediction. & Outperformed traditional models for long-term satellite clock bias prediction.\\
\hline
Yun et al. \cite{yun_practical_2022} & Practical approach using dual-frequency GNSS measurements to improve smartphone position accuracy. & Overcoming limitations of smartphones and leveraging dual-frequency measurements for quality monitoring.\\
\hline
Mendonca et al. \cite{mendonca_rtk_2022} & Genetic Algorithm-based machine learning classifier for validating ambiguity terms in RTK positioning. & Improved classification performance compared to traditional ratio test.\\
\hline
Lacambre et al. \cite{lacambre_optimizing_2022} & Methodology incorporating ML methods for optimizing and qualifying new RTK GNSS algorithms. & Improved real-world positioning performance through outlier detection and reference comparison.\\
\hline
Sun et al. \cite{sun_improving_2021} & GBDT-based approach for correcting GPS positioning errors caused by MP and NLOS signals. & Significant improvement in 3D positioning accuracy compared to conventional methods.\\
\hline
Mohanty et al. \cite{meold, menew} & Graph Convolution Network and Kalman Filter. & Improved accuracy compared to model-based and learning-based methods.\\
\hline
Zhao et al. \cite{gnnrnn} & Graph Neural Network combined with Reinforcement Learning. & 26 \% improvement in urban datasets; 10 \% improvement in semi-urban.\\
\hline
\end{tabular}
\end{table}

\textbf{Recurrent Neural Networks:} Kim et al. \cite{kim_deep_2019} used LSTM-based recurrent neural networks to enhance accuracy and stability in autonomous vehicle navigation. Yang et al. \cite{yang_build_2019} furthered this effort by developing an LSTM RNN model tailored for real-time prediction of GPS positioning errors. Other works, such as those by Thomas et al. \cite{thomas_position_2021} and Zhou et al. \cite{zhou_least-squares_2017}, explored ML-based post-processing techniques for improving position accuracy in autonomous vehicle applications and GNSS navigation integrated with Kalman filtering, respectively.
Liu et al. \cite{liu_vehicular_2019} used LSTM-based prediction to enhance the accuracy of GPS in vehicular navigation.

\textbf{Other Methods and Applications:} The research by Neri et al. \cite{neri_machine_2020} is tailored specifically for the rail domain. They aim to enhance the accuracy and reliability of train positioning systems by combining classical observables with advanced RAIM techniques. 
A unique approach is introduced in \cite{rl_adapt} to enhance high-precision GNSS positioning in dynamic urban terrains using a deep reinforcement learning framework.
Random Forest was explored in~\cite{new_random_forest} along with conformal prediction to learn positioning errors and integrity intervals with $99.999 \%$ confidence.

\section{Other Use Cases}
In various GNSS applications, different ML methods have been employed to go beyond improving the receiver's positioning performance. 
While these applications are not the main focus of our survey paper, we provide a concise overview of some existing works.

\textbf{GNSS Augmentation Systems and Carrier Phase Measurements:} The authors in \cite{car_anomaly} presented an anomaly detection algorithm tailored for carrier phase measurements in GNSS augmentation systems. Targeting safety-critical applications like autonomous vehicles, their machine learning-based approach estimates standard deviations of residual errors. This enables continuous fault monitoring even with single-frequency measurements, and the real-world tests validate the method's efficiency.

\textbf{Direction-of-Arrival (DOA) Estimation:} In \cite{vul_gps1}, a novel method for DOA estimation is introduced. Unlike conventional neural network-based approaches, this method addresses real-world array imperfections. A Transformer-based calibration network (TCN) models these imperfections at the antenna level, utilizing global and long-term properties of array errors. Experiments indicate superiority over traditional techniques, especially under amplitude and phase deviations and antenna position perturbations.

\textbf{Velocity and Acceleration Measurements:} The study in \cite{chang_precise_2020} analyzed the performance of a stand-alone GNSS receiver after incorporating sparse kernel learning. This AI-based solution improves accuracy in measuring velocity and acceleration, paving the way for applications in vehicle dynamics analysis and geodetic monitoring.

\textbf{Ionospheric Prediction and Total Electron Content (TEC) Variations:} The work in \cite{chen_real-time_2023} presented a real-time ionosphere prediction model using LSTM, utilizing International GNSS Service products to estimate and correct ionospheric delays. 
In \cite{tec_1}, LSTM and Transformer networks are used to predict TEC variations, showing performance enhancements over traditional methods. Additionally, a spatiotemporal Graph Neural Network, coupled with transformers, is utilized in \cite{vtec1} for predicting VTEC maps. Graph nodes in this approach symbolize pixels holding VTEC values, while edges are determined by inter-node distances. Another deep learning-based system in \cite{ionodelay1} is used to forecast TEC maps for South America's ionosphere. Finally, the global TEC map prediction framework in \cite{tec2} compared LSTM and Transformer networks, illustrating the superiority of the suggested networks over IGS rapid products.

\textbf{Earthquake Detection and Environmental Characterization:} The research in \cite{dittmann_supervised_2022} used supervised machine learning for analyzing GNSS velocities tied to earthquake-strong motion signals. The models, trained on datasets of strong motion events, offer increased accuracy in seismic activity detection. On a related note, \cite{dovis_opportunistic_2020} used ML to exploit GNSS signals for environmental characterization. Through this approach, they extract significant environmental data, facilitating applications in climate studies, precision agriculture, and environmental monitoring.

\textbf{GNSS Functional Safety and Satellite Orbit Predictions:} Ensuring GNSS functional safety is at the forefront of \cite{gogliettino_machine_2019}. By leveraging machine learning, this research addressed potential safety hazards in GNSS systems. Orbit prediction for Low Earth Orbit satellites, as explored in \cite{haidar-ahmad_hybrid_2022} combined analytical models with ML techniques to predict orbits, respectively. Another model, presented in \cite{orbit2}, employed a transformer deep learning framework for satellite orbit correction prediction, outclassing existing prediction methods.

\textbf{Satellite Selection, Interference Detection, and Data Fusion:} The deep learning network in \cite{huang_satellite_2018} performed optimal satellite selection in GNSS positioning. By intelligently accounting for factors like signal quality, satellite geometry, and user demands, this model promises enhanced positioning accuracy. Works like \cite{kazemi_enhancing_2020} and \cite{liu_gnss_2021} used ML for GNSS interference detection and classification. Finally, \cite{navarro_data_2021} and \cite{navarro_data-intensive_2021} used ML techniques for innovative GNSS science applications, unlocking potential in atmospheric sensing and climate studies.

\textbf{Spoofing Detection and Signal Security:} Several contributions target GNSS spoofing detection. Research like \cite{aissou_tree-based_2021}, \cite{bose_gps_2022}, \cite{feng_gnss_2022}, and \cite{jullian_deep_2022} employed diverse machine learning techniques ranging from tree-based models to deep learning for effective spoofing detection. While \cite{mehr_detection_2022} and \cite{pardhasaradhi_machine_2022} focused on jamming detection and measurement association in spoofing environments, \cite{semanjski_use_2019} used supervised ML for detecting GNSS signal spoofing, further showcasing the significance of MLin ensuring GNSS signal integrity and security.

\section{Limitations of Discussed Methods and Potential Solutions}

Although the methods discussed for enhancing GNSS positioning through ML demonstrate promising results, it is essential to consider their limitations. Additionally, we provide potential solutions to directly address these limitations which include the following:

\begin{enumerate}
    \item \textbf{Data Dependency:} Many of the ML-based methods rely heavily on the availability of large and diverse datasets for training. While data-driven approaches have demonstrated success in improving GNSS positioning, gathering and maintaining such datasets can be challenging. Insufficient or biased training data may limit the generalizability and effectiveness of the ML models. Furthermore, collecting data for specific environments or rare scenarios may be time-consuming and resource-intensive. Adequate data collection efforts and quality control measures are necessary to ensure the reliability of ML models.

    \item \textbf{Computational Requirements:} ML models, especially deep learning models, often require significant computational resources for training. Training deep learning models on large datasets can be computationally intensive and time-consuming. Deploying these models in resource-constrained environments, such as embedded systems or low-power devices, may pose challenges. Developing lightweight ML models or exploring alternative architectures that strike a balance between computational efficiency and accuracy is essential for practical deployment.

    \item \textbf{Generalizability to Unseen Scenarios:} While ML models trained on extensive datasets can exhibit impressive performance in controlled test environments, their generalizability to unseen or evolving scenarios remains a concern. Changes in satellite constellations, emerging technologies, or novel interference sources may require model retraining or adaptation. Ensuring the long-term effectiveness and adaptability of these models in dynamic GNSS environments requires continuous monitoring, updating, and reevaluation of the models.

    \item \textbf{Dependency on GNSS Signal Availability:} ML models designed to improve GNSS positioning heavily rely on the availability of GNSS signals. However, there are instances when GNSS signals may be temporarily unavailable or degraded due to signal blockage, jamming, or interference. In such cases, the performance of ML models that rely solely on GNSS inputs may be limited. Developing hybrid positioning approaches that combine GNSS with other sensors, such as inertial sensors or environmental context data, can help mitigate this limitation and provide robust positioning solutions.

    \item \textbf{Lack of Standardization:} The field of ML for improving GNSS positioning is still evolving, and there is a lack of standardized methodologies, evaluation metrics, and benchmark datasets. The absence of standards makes it challenging to compare and replicate results across different studies. Developing standardized evaluation frameworks, sharing benchmark datasets, and promoting reproducibility are essential for advancing the field and enabling meaningful comparisons between different AI-based methods.

    \item \textbf{Integration Complexity:} Integrating ML-based algorithms into existing GNSS positioning systems can be complex and may require modifications to the system architecture or hardware. Compatibility issues, system interoperability, and deployment challenges need to be addressed to ensure seamless integration and practical implementation of ML techniques. Collaborative efforts among ML researchers, GNSS experts, and industry stakeholders are necessary to overcome integration barriers and facilitate the adoption of ML in real-world GNSS positioning applications.

    \item \textbf{Cost and Scalability:} The implementation of ML-based methods for improving GNSS positioning may involve initial investment costs, including infrastructure, computational resources, and expertise. The scalability of ML models to handle large-scale positioning systems and accommodate increasing data volumes may also
pose challenges. Ensuring cost-effective solutions and scalability is crucial for the practical adoption of ML techniques in GNSS positioning. Exploring cloud-based solutions, distributed computing, or edge computing approaches can help address scalability concerns and optimize resource utilization.
    
\end{enumerate}

\section{Promising Opportunities}

Several promising opportunities arise for the application of ML techniques to enhance GNSS positioning systems.
We discuss some key opportunities below.

\begin{enumerate}
    \item \textbf{Integration with Other Sensor Modalities:} ML techniques offer opportunities for seamless integration of GNSS data with other sensor modalities, such as IMUs, odometers, or digital maps. 
By leveraging the complementary information from different sensor modalities, ML-based integration methods can overcome limitations associated with individual sensors and provide more reliable and accurate positioning solutions.
GNNs can also be employed to integrate GNSS measurements with data from other sensors, such as LiDAR or camera sensors. By representing the sensor data as a graph structure and leveraging GNNs, the models can capture the complex relationships and dependencies between different sensor modalities. This integration allows for more comprehensive and accurate positioning solutions, especially in scenarios where GNSS signals may be affected by obstructions or limitations.

\item   \textbf{Adaptive Algorithms for Dynamic Environments:} ML algorithms can adapt and learn from dynamic environments, allowing for real-time adjustments in GNSS positioning. These algorithms can continuously analyze and update models based on changing environmental conditions, satellite availability, or user dynamics. By considering factors such as satellite constellation health, signal quality, and user motion patterns, ML-based algorithms can dynamically optimize positioning solutions to provide accurate and reliable results.

\item \textbf{Crowd-Sourced Positioning}: ML techniques can harness the power of crowd-sourced data to enhance GNSS positioning accuracy. By collecting positioning data from a large number of users and applying machine learning algorithms, patterns, and trends can be extracted to improve the overall accuracy of positioning solutions. This approach can be especially beneficial in areas with limited GNSS coverage or challenging signal conditions, as it relies on collective data contributions to overcome individual limitations.

\item \textbf{Transfer Learning for Cross-Domain Positioning}: Transfer learning techniques can be applied to leverage knowledge gained from one GNSS domain to another with limited data. 
Such an approach can save data collection efforts and enhance the performance of GNSS systems in underrepresented domains.

\item \textbf{Meta-Learning for Adaptive GNSS Algorithms:} Meta-learning algorithms can be used to learn the optimal algorithmic configurations for GNSS positioning based on historical performance data. By training a meta-learning model on a variety of datasets, it can learn which algorithms work best under different conditions and adaptively select or combine them to achieve optimal positioning accuracy. This adaptive approach allows GNSS systems to continuously improve their performance and adapt to changing environments.

\item \textbf{Generative Adversarial Networks (GANs) for Data Augmentation:} GANs can be used to generate synthetic GNSS data that closely resemble real-world measurements. By training a GAN on a large dataset of GNSS observations, it can learn the underlying distribution of the data and generate additional samples. These synthetic samples can be used to augment the training data for GNSS positioning algorithms, thereby improving their performance, especially in scenarios with limited training data.

\item \textbf{Uncertainty Estimation using Bayesian Neural Networks:} Utilize Bayesian neural networks to estimate uncertainty in GNSS positioning solutions, providing confidence intervals and probabilistic measures of accuracy for better decision-making in critical applications.

\item \textbf{Federated Learning:} Employ federated learning approaches to train positioning models collaboratively across multiple devices or users, ensuring privacy while improving the accuracy and robustness of GNSS positioning.

\item \textbf{Edge Computing for Real-time GNSS Processing:} Utilize edge computing architectures to perform real-time GNSS data processing and positioning calculations at the network edge, reducing latency and enabling faster and more responsive positioning solutions.

\end{enumerate}

\section{Conclusion}
In conclusion, this survey paper has explored the application of ML methods for GNSS-based positioning. 
The paper has provided a comprehensive overview of various ML techniques and their relevance to different aspects of GNSS positioning. It has covered topics such as signal analysis and classification, environmental context recognition, anomaly detection, multi-sensor integration, prediction and forecasting, accuracy enhancement, and position error modeling. Additionally, the paper has discussed other notable applications of ML in GNSS and has identified the limitations and challenges associated with these methods. The survey concludes by highlighting potential areas for future research and development in the field of ML-based GNSS positioning. 
Overall, this survey contributes to a deeper understanding of the role of ML in improving GNSS positioning and provides valuable insights for researchers and practitioners in the field.


\bmhead{List of Abbreviations}
\begin{itemize}
    \item \textbf{GNSS} - Global Navigation Satellite Systems
    \item \textbf{ML} - Machine Learning
    \item \textbf{GPS} - Global Positioning System
    \item \textbf{NLOS} - Non-Line-of-Sight
    \item \textbf{MP} - MP
    \item \textbf{WLS} - Weighted Least Squares
    \item \textbf{RTK} - Real-time kinematic
    \item \textbf{PPP} - Precise Point Positioning
    \item \textbf{SNR} - Signal-to-noise ratio
    \item \textbf{SVMs} - Support Vector Machines
    \item \textbf{CNNs} - Convolutional Neural Networks
    \item \textbf{SVR} - Support Vector Regression
    \item \textbf{DNN} - Deep Neural Network
    \item \textbf{RNN} - Recurrent Neural Network
     \item \textbf{KNN} - K-Nearest Neighbors
    \item \textbf{LSTM} - Long Short-Term Memory
    \item \textbf{MLP} - Multilayer Perceptron
    \item \textbf{RBFNN} - Radial Basis Function Neural Network
    \item \textbf{GNN} - Graph Neural Network
    \item \textbf{GraphSAGE} - Graph Sample and Aggregation
    \item \textbf{GBDT} - Gradient Boosting Decision Tree
    \item \textbf{XGBoost} - Extreme Gradient Boosting
    \item \textbf{RL} - Reinforcement Learning
    \item \textbf{DQN} - Deep Q-Networks
    \item \textbf{PPO} - Proximal Policy Optimization
    \item \textbf{A2C} - Advantage Actor-Critic
    \item \textbf{VAE} - Variational Autoencoders
    \item \textbf{KL} - Kullback-Leibler
    \item \textbf{LOS} - Line-of-sight
    \item \textbf{SVM} - Support Vector Machine
    \item \textbf{NN} - Neural Network
    \item \textbf{RBF SVM} - Radial Basis Function Support Vector Machines
    \item \textbf{ML} - Machine Learning
    \item \textbf{C/N0} - Carrier-to-Noise Ratio
    \item \textbf{NavIC} - Navigation with Indian Constellation
    \item \textbf{L5} - Likely referring to a specific GNSS signal frequency band
    \item \textbf{DLL} - Delay-Locked Loop
    \item \textbf{ASIC} - Application-Specific Integrated Circuit
    \item \textbf{BDS} - BeiDou Navigation Satellite System
    \item \textbf{GRU} - Gated Recurrent Unit
    \item \textbf{GTNN} - Graph Transformer Neural Network
    \item \textbf{IMU} - Inertial Measurement Unit
    \item \textbf{RAIM} - Receiver Autonomous Integrity Monitoring
    \item \textbf{TDNN} - Time-delayed
Neural Network
    \item \textbf{PFRNN} - Particle Filter Recurrent Neural Network
Neural Network
\end{itemize}




\bmhead{Acknowledgments}
We would like to acknowledge Derek Knowles and Asta Wu for providing feedback on the paper.


\section*{Declarations}

\begin{itemize}
\item Funding: Not applicable
\item Conflict of interest/Competing interests: The authors declare that they have no competing interests.
\item Ethics approval: Not applicable
\item Consent to participate: Not applicable
\item Consent for publication: Yes
\item Availability of data and materials: Data sharing is not applicable to this article as no datasets were generated or analyzed during the current study.
\item Code availability: Not applicable
\item Authors' contributions: AM wrote the first draft of the manuscript and conducted the survey. AM and GG finalized the manuscript write-up, proofread, and checked the technical correctness of the manuscript. 
All authors read and approved the final manuscript.
\end{itemize}

\bibliography{sn-bibliography}

\end{document}